\documentclass[journal]{IEEEtran}
\usepackage[english]{babel}

\ifCLASSINFOpdf
\else
\fi
\usepackage[T1]{fontenc}
\usepackage[latin9]{inputenc}
\usepackage{xcolor}
\usepackage{amsmath}
\usepackage{amssymb}
\usepackage{color}
\usepackage{graphicx}

\usepackage{wasysym}
\usepackage{wasysym}
\PassOptionsToPackage{normalem}{ulem}
\usepackage{ulem}
\usepackage{float}
\usepackage{ifsym}
\usepackage[colorlinks,
linkcolor=teal,
anchorcolor=teal,
citecolor=teal]{hyperref} 

\floatstyle{ruled}
\newfloat{algorithm}{tbp}{loa}
\providecommand{\algorithmname}{Algorithm}
\floatname{algorithm}{\protect\algorithmname}

\newtheorem{thm}{Theorem}
\newtheorem{problem}{Problem}
\newtheorem{rem}{Remark}
\newtheorem{lem}{Lemma}
\newtheorem{prop}{Proposition}
\newtheorem{example}{Example}

\begin{document}
%
\title{Logic-based switching finite-time stabilization with applications in mechatronic systems}
%
%
%

\author{Shiqi~Zheng,~\IEEEmembership{Senior Memeber,~IEEE},    
	Shihao~Wang, 
     Xiang Chen,
     and Yuanlong Xie,~\IEEEmembership{Memeber,~IEEE}    
\thanks{The work was supported by the National Natural Science Foundation
	of China (Grant No. 52105019, No. 61703376), Open Research Project of Zhejiang Lab  (Grant No. 2022NB0AB03) and 111 project (Grant No. B17040). (\emph{Corresponding author: Shiqi Zheng and Yuanlong Xie)}}
\thanks{S. Zheng, S. Wang and X. Chen are with the School of Automation, China University of Geosciences, and
Hubei key Laboratory of Advanced Control and Intelligent Automation for Complex Systems, Wuhan, China (e-mail: zhengshiqi@cug.edu.cn. tel: +86-13554039454).

Yuanlong Xie is with the School of Mechanical Science and Engineering, Huazhong University of Science and Technology, Wuhan 430074, China (e-mail: yuanlongxie@hust.edu.cn).}
}

%
%

\markboth{IEEE Transactions on Cybernetics}%
{Logic-based switching finite-time stabilization with applications in mechanical systems}
%



\maketitle

\begin{abstract}
This paper investigates the finite time stabilization problem for
a class of nonlinear systems with unknown control directions and unstructured
uncertainties. The unstructured uncertainties indicate that not only
the parameters but also the structure of the system nonlinearities
are uncertain. A new adaptive control method is proposed for the considered
system. Logic-based switching rule is utilized to tune the controller
parameters online to stabilize the system in finite time. Different
from the existing adaptive controllers for structured/parametric uncertainties,
a new switching barrier Lyapunov method and supervisory functions
are introduced to overcome the obstacles caused by unstructured uncertainties
and unknown control directions. Both simulations and experiments are conducted
on mechatronic systems to verify the effectiveness of the proposed methods.
\end{abstract}


\section{Introduction}

\subsection{Background and motivations}

Finite time stabilization problem has attracted increasing attention
in the past few years. Finite time stabilization means that by designing
a proper feedback controller, all the states of the closed loop systems
will become exact zero after finite time \cite{key-33-1}. However, for asymptotic stabilization, the states will
converge to zero in an infinite time. Lots of works \cite{key-4-1,key-3s,key-1aa} have shown that finite time control has some promising
features in contrast with asymptotic control. These may lie in: 1)
Faster convergence rate and higher precision; 2) Possibility to decouple
the stabilization problem from other control objectives \cite{key-1}. 

Many interesting results have been obtained for finite time control.
The works of \cite{key-33-1}
finite time output feedback stabilization for strict feedback nonlinear
systems. \cite{key-34-1} have extended
the finite time control to high order stochastic nonlinear systems.
A time-varying feedback method is proposed in \cite{key-18a} to achieve prescribed finite time
control performance. Namely, the finite convergence time can be determined
\textit{a prior} and is independent of the initial conditions. Recently,
the finite time control problem has been investigated for multi-agent
and networked systems \cite{key-4111}. Moreover,
several real practical applications, such as robot manipulators \cite{key-77} and servo motor systems \cite{key-88}, \cite{key-35} have
been considered for finite time control.

Unknown control directions are often encountered in real engineering
world. It means that the sign of control coefficient is unknown. This
will bring difficulties to the controller design because a control
effort with wrong direction can drive the states away from the equilibrium
point. Nussbaum-gain technique, which was originally introduced in
\cite{key-18}, is a common way to handle unknown
control direction. Plentiful works \cite{key-4,key-21,key-20,key-2,key-31} have been done on the control of nonlinear systems
by incorporating Nussbaum-gain function. Nevertheless, as discussed
in \cite{key-25} and \cite{key-26}, the Nussbaum-gain technique could only achieve
asymptotic stability because the constructed Lyapunov function cannot
be negative definite.

In fact, there are very few works concentrating on finite time stabilization
of nonlinear systems with unknown control directions. Lately, in the
framework of backstepping method, a new adaptive control strategy is proposed in
\cite{key-25,key-26} and \cite{key-29} to solve this
problem. The idea of the method is to adopt a logic-based switching
rule to tune the controller parameters online according to a well-defined
supervisory function. Finite time stability can then be achieved despite
unknown control directions. 

The aforementioned works \cite{key-25,key-29}, however, only consider the finite time stabilization
problem for nonlinear systems suffering from structured/parametric
uncertainties. This means that the structures of the nonlinear uncertain
functions are available, but contain some unknown parameters. The
structured uncertainties are mainly used to describe the parameter
variations in the systems. However, the nonlinear uncertainties are
often very complicated in practical systems. Hence, it may be difficult
or impossible to obtain the exact form of the uncertainties, and express
the uncertainties in a parametric way. This class of uncertainties
is often referred to as unstructured/nonparametric uncertainties,
which can represent those unknown nonlinearities caused by complex
system dynamics and modeling errors. Therefore, a natural question
arises:

\textit{How to solve the finite time stabilization problem for nonlinear
	systems with unknown control directions and unstructured uncertainties? }

To the best of our knowledge, little effort has been made to answer
the above issue. The main challenges may lie in the following aspects: 

1) Due to the structure of the nonlinearities is uncertain, the nonlinearities
cannot be parameterized. Hence, it is difficult to directly extend
the adaptive control scheme presented in \cite{key-25} and \cite{key-4a} to solve
the above problem. Consequently, the design procedures become involved.

2) As previously mentioned, a logic-based switching mechanism has
to be adopted to achieve finite time stability due to the possible
limitations of the Nussbaum-gain technique.\textcolor{black}{{} Therefore,
	the entire closed-loop system will exhibit }\textit{\textcolor{black}{hybrid}}\textcolor{black}{{}
	feature, which introduces difficulties to the controller design and
	stability analysis.}


\subsection{Contributions}

Motivated by the above thought, this paper focuses on the finite time
stabilization problem for a class of nonlinear systems with unknown
control directions and unstructured uncertainties. The contributions
are mainly in the following aspects.\smallskip{}

A new switching adaptive control method is proposed for the
considered system. Logic-based switching rule is used to tune the
controller parameters online. The proposed method includes two novel
techniques:
\begin{itemize}
	\item Novel \textit{switching barrier Lyapunov functions} are constructed
	for the controller design. The barrier will switch according to the
	logic-based switching rule (see Remark \ref{rem:(Switching-barrier-Lyapunov}). 
	\item By designing some special auxiliary systems, \textit{new supervisory
		functions} are presented to guide the logic-based switching (see Remark
	\ref{rem:(New-supervisory-functions)}).
\end{itemize}
Based on the above two new techniques, the states of the \emph{hybrid} closed-loop
systems will be constrained in a compact set despite multiple unknown
control directions. Then, constant bounds will be obtained for the
unstructured uncertainties. This contributes to the feasibility of
the adaptive control scheme such that all the states will reach exact
zero in finite time.\smallskip{}
%

Moreover, the proposed methods have some promising features, such
as fast convergence speed, small control overshoot, low complexity
and strong robustness to unknown control directions.  

\subsection{Organizations}

The organization of the paper is as follows. Problem formulation and
preliminaries are presented in Section II. Section III concentrates on
the finite time controller design.  Simulations and experiments are conducted in Sections IV-V.
Section VI presents some discussions and conclusions. Proofs 
are provided in Appendices.

\textit{Notations.} Given a real number $x$ and a positive constant
$\alpha=\frac{p}{q}$ where $p,q$ are coprime. If $p$ is a positive
odd integer, then $x^{\alpha}=\mathrm{sign}(x)|x|^{\alpha}.$ If $p$
is a positive even integer, then $x^{\alpha}=|x|^{\alpha}.$ \textcolor{black}{Let $a,b\in\mathbb{R}$, then $a\triangleq b$ means $a$ is defined as $b$. $a:=b$ means $a$ is set as $b$, which is used in algorithm.}

\section{Preliminaries and problem formulation}

\subsection{Problem formulation}

Consider the following system
\begin{align}
\dot{x}_{i} & =h_{i}(\overline{x}_{i})x_{i+1}+f_{i}(\overline{x}_{i}),\thinspace i=1,2,...,n-1\nonumber \\
\dot{x}_{n} & =h_{n}(\overline{x}_{n})u+f_{n}(\overline{x}_{n}),\label{eq:1-1}\\
y & =x_{1}\nonumber 
\end{align}
where $\overline{x}_{i}=(x_{1},x_{2},...,x_{i})^{\mathrm{T}}\in\mathbb{R}^{i},\thinspace i=1,2,...,n$
are the system states, $y$ is the system output. $f_{i}(\overline{x}_{i}),h_{i}(\overline{x}_{i})(i=1,...,n)$
are all \textit{unknown} continuously differentiable nonlinear functions.
$f_{i}(\overline{x}_{i})$ represents the system nonlinearities and
uncertainties such that $f_{i}(0,0,...,0)\equiv0$. $h_{i}(\overline{x}_{i})$
are the control gains such that their signs are \textit{unknown} and
satisfy $|h_{i}(\overline{x}_{i})|>0$. $u$ denotes the control input. 
\begin{rem}\emph{\textcolor{black}{(More general systems)}}
	Note that system (\ref{eq:1-1}) is more general than the existing
	works \cite{key-35,key-25,key-26,key-30,key-15-1} due to the following reasons:
	
	1) $f_{i}(\overline{x}_{i}),h_{i}(\overline{x}_{i})(i=1,2,...,n)$
	represent unstructured uncertainties, \textit{i.e.}, not only parameters
	in $f_{i}(\overline{x}_{i}),h_{i}(\overline{x}_{i})$ but also the
	form of $f_{i}(\overline{x}_{i}),h_{i}(\overline{x}_{i})$ are uncertain.
	In fact, the continuously differentiable nonlinearities $f_{i}(\overline{x}_{i}),h_{i}(\overline{x}_{i})$
	only need to satisfy $f_{i}(0,0,...,0)\equiv0,|h_{i}(\overline{x}_{i})|>0$.
	This is much more general than structured uncertainties in \cite{key-25,key-29} and \cite{key-4a}. In these references, $f_{i}(\overline{x}_{i}),h_{i}(\overline{x}_{i})$
	need to satisfy $|f_{i}(\overline{x}_{i})|\leq(|x_{1}|+\cdots+|x_{i}|)\psi_{i}(\overline{x}_{i},\theta)$
	and $0<\underline{h}_{i}\leq|h_{i}(\overline{x}_{i})|\leq\overline{h}_{i}(\overline{x}_{i},\theta)$
	where $\psi_{i}(\overline{x}_{i},\theta)$ and $\overline{h}_{i}(\overline{x}_{i},\theta)$
	are known smooth functions, $\theta,\underline{h}_{i}$ are unknown
	parameters. 
	
	2) System (\ref{eq:1-1}) contains multiple unknown control directions,
	\textit{i.e.}, for $\forall i=1,2,..,n$, the sign of $h_{i}(\overline{x}_{i})$
	is unknown. Moreover, compared with \cite{key-2}
	and \cite{key-32}, $|h_{i}(\overline{x}_{i})|$
	only needs to be larger than zero, not a positive constant. 
	
	According to the above analysis, we can see that very little information
	is needed for $f_{i}(\overline{x}_{i}),h_{i}(\overline{x}_{i})$.
	This will bring many difficulties to the controller design. In addition,
	note that by unstructured/nonparametric uncertainties, it means that
	it is difficult to obtain the \textit{exact} form of the uncertainties,
	and the uncertain functions cannot be parameterized by unknown parameters.
	However, some crude information of the uncertainties may need to be
	known. For instance, the nonlinear function $f_{i}(\overline{x}_{i})$
	needs to satisfy $f_{i}(0,0,...,0)\equiv0$. Yet, we can see the structures
	of the nonlinearities are still uncertain because many kinds of nonlinear
	functions satisfy $f_{i}(0,0,...,0)\equiv0$. ~~\Square{}
\end{rem}

Now, we are ready to describe the finite time stabilization problem. 
\begin{problem}
	\textit{(Finite time stabilization problem) }Develop an adaptive controller
	$u(t)$ for system (\ref{eq:1-1}) such that
	
	1) All the control signals in the closed loop system are bounded,
	and;
	
	2) All the states will converge to zero in finite time, $i.e.$, there
	exists a finite time $T$ such that $x_{i}(t)\rightarrow0(i=1,2,...,n)$
	as $t\rightarrow T^{-}$.
\end{problem}

\subsection{Technical lemmas}

Some useful lemmas will be presented, which will be used in the controller
design.
\begin{lem}
	\label{lem:Q1} (\cite{key-15-1}) Consider
	the following Young's inequality
	\[
	|x|^{a}|y|^{b}\leq\frac{a}{a+b}\zeta(x,y)|x|^{a+b}+\frac{b}{a+b}\zeta^{-a/b}(x,y)|y|^{a+b}
	\]
	where $x,y\in\mathbb{R}$, $a,b$ are positive constants, $\zeta(x,y)>0$
	is any real valued function.
\end{lem}
\begin{lem}
	\label{lem:Q2} (\cite{key-15-1,key-28}) Given a real constant $p\geq1$ being a ratio of
	two odd integers and real numbers $x,y,z_{i}(i=1,2,...,n)$, we have:
	\[
	|x-y|^{p}\leq2^{p-1}|x^{p}-y^{p}|;
	\]
	\[
	|x^{1/p}-y^{1/p}|\leq2^{1-1/p}|x-y|^{1/p};
	\]
	\[
	\left(\sum_{i=1}^{n}|z_{i}|\right)^{1/p}\leq\sum_{i=1}^{n}|z_{i}|^{1/p}\leq n^{1-1/p}\left(\sum_{i=1}^{n}|z_{i}|\right)^{1/p}.
	\]
\end{lem}
\begin{lem}
	\label{lem:A1}(\textcolor{teal}{Lemma 11.1 in} \cite{key-1a,key-4a-1}) Given a continuously differentiable
	nonlinear function $f(\overline{x}_{n}):\mathbb{R}^{n}\rightarrow\mathbb{R}$
	where $\overline{x}_{n}=(x_{1},x_{2},...,x_{n})^{\mathrm{T}}\in\mathbb{R}^{n}$
	and $f(0,0,...,0)\equiv0$,  there exists a non-negative smooth function
	\textup{$\psi(\overline{x}_{n}):\mathbb{R}^{n}\rightarrow\mathbb{R}$}
	such that
	\begin{equation}
	|f(\overline{x}_{n})|\leq(|x_{1}|+\cdots+|x_{n}|)\psi(\overline{x}_{n}).\label{eq:a}
	\end{equation}
\end{lem}
\begin{lem}
	\label{lem:C1}Given four time-varying continuous functions $x(t),y(t),a(t),b(t):[0,+\infty)\rightarrow\mathbb{R}$
	such that 
	\begin{align}
	\dot{x}(t) & =-a(t)x^{\gamma}(t)+b(t),\label{eq:C1}\\
	\dot{y}(t) & \leq-a(t)y^{\gamma}(t)+b(t)\label{eq:C2}
	\end{align}
	for $\forall t\in[t_{0},t_{1})\subseteq[0,+\infty)$ where $x(t_{0})=y(t_{0})+\varepsilon$.
	$\varepsilon\geq0$ and $0<\gamma\leq1$ are constants with $\gamma$
	being a ratio of odd integers, $a(t)>0$ on $[0,+\infty)$. Then,
	$x(t)\geq y(t)$ for $\forall t\in[t_{0},t_{1})$.
\end{lem}
\begin{IEEEproof}
	Please see Appendix A for detailed proof.
\end{IEEEproof}

\section{Finite time stabilization}

\textcolor{black}{This section will focus on the finite time stabilization
	of system (\ref{eq:1-1}). It is divided into three parts. In Section
	III-A, we will mainly present the controller structure, which contains
	some adaptive parameters. Section III-B will focus on the logic-based
	switching rule for tuning the adaptive parameters. Main result and
	stability analysis for the hybrid closed-loop system will be given
	in Section III-C. }

\subsection{\textcolor{black}{Controller design}}

It is noted that the high order system (\ref{eq:1-1}) can be regarded
as a cascade of $n$ first order subsystems. The controller design
for this class of system is inspired by backstepping method \cite{key-4a}. The controller $u$ is recursively
determined by the following equations:
\begin{align}
s_{1} & \triangleq x_{1},\label{eq:n1}\\
x_{i+1}^{*} & \triangleq\hat{\Theta}_{i}(t)\left[-K_{i}s_{i}^{q_{i+1}}-\frac{U_{i}s_{i}^{q_{i+1}}}{(\hat{\chi}_{i}^{2}(t)-s_{i}^{2})^{1+2\alpha}}\right],\label{eq:n2}\\
s_{i+1} & \triangleq x_{i+1}^{1/q_{i+1}}-\left(x_{i+1}^{*}\right)^{1/q_{i+1}}(i=1,2,...,n-1),\label{eq:n3}\\
x_{n+1}^{*} & \triangleq u\triangleq\hat{\Theta}_{n}(t)\left[-K_{n}s_{n}^{q_{n+1}}-\frac{U_{n}s_{n}^{q_{n+1}}}{(\hat{\chi}_{n}^{2}(t)-s_{n}^{2})^{1+2\alpha}}\right].\label{eq:n4}
\end{align}
$x_{i+1}^{*}(i=1,2,...,n-1)$ are the virtual control efforts for
the first order subsystem $\dot{x}_{i}=h_{i}(\overline{x}_{i})x_{i+1}+f_{i}(\overline{x}_{i})$
in (\ref{eq:1-1}). $s_{i}(i=1,2,...,n)$ are the virtual control
errors which will be regulated to zero in finite time. 

The design parameters $q_{i+1}(i=1,2,...,n)$ are determined by $q_{i+1}=\alpha-1+q_{i}$
where $\alpha\in(\frac{1}{2},1)$ is a ratio of odd integers and $q_{1}\triangleq1$.
These parameters are accounting for the power of Lyapunov function
which is the key for the finite time stabilization.\textcolor{black}{{}
}$K_{i},U_{i}(i=1,...,n)$ are positive design parameters, $\hat{\chi}_{i}(t)$
and $\hat{\Theta}_{i}(t)$ are adaptive parameters explained as follows:

\textcolor{black}{$\hat{\chi}_{i}(t):[0,+\infty)\rightarrow\mathbb{R}(i=1,2,...,n)$
	are piecewise constant signals. It will be updated according to the
	logic-based switching rule in Section 3.2. $\hat{\chi}_{i}(t)(i=1,...,n)$
	are used to constrain all the virtual control errors $s_{i}$ and
	states in a compact set, which is beneficial for dealing with the
	unstructured uncertainties.}

\textcolor{black}{$\hat{\Theta}_{i}(t):[0,+\infty)\rightarrow\mathbb{R}$
	is determined by a switching signal $\sigma_{i}(t)$.
	\begin{equation}
	\hat{\Theta}_{i}(t)=(-1)^{\sigma_{i}(t)}\theta_{i}(\sigma_{i}(t))\label{eq:8-1}
	\end{equation}
	where $\sigma_{i}(t):[0,+\infty)\rightarrow\mathbb{N}$ is also a
	piecewise constant signal, $\theta_{i}(\sigma_{i}):\mathbb{N}\rightarrow\mathbb{R}$
	is an increasing function with respect to $\sigma_{i}$ such that
	$\theta_{i}(0)>0$ and $\theta_{i}(\sigma_{i})\rightarrow+\infty$
	as $\sigma_{i}\rightarrow+\infty$. A typical example of $\theta_{i}(\sigma_{i})$
	is $\theta_{i}(0)=1,\theta_{i}(1)=2,\theta_{i}(2)=3,...$ The idea
	of  the tuning rule (\ref{eq:8-1}) for $\hat{\Theta}_{i}$ is that
	by changing its sign repeatedly, one may expect to find a correct
	control direction, $i.e.,$ dealing with the unknown sign of $h_{i}(\overline{x}_{i})$
	in (\ref{eq:1-1}). The detail switching rule for $\sigma_{i},\hat{\Theta}_{i}$
	is also given in Section III-B. }

\textcolor{black}{Next, a $n$ step Lyapunov functions analysis will
	be given according to $x_{i+1}^{*}(i=1,2,...,n)$. This analysis will
	be helpful for understanding the controller design idea and results
	in Sections III-B and III-C. It should be noted that all the adaptive parameters
	$\hat{\chi}_{i},\hat{\Theta}_{i}$ will be assumed to be constants
	in the following results. This will become clear in Section III-C.}

\textit{Step 1}. Consider the following Lyapunov function 
\begin{equation}
V_{1}=\frac{1}{2}\ln\left(\frac{\hat{\chi}_{1}^{2}(t)}{\hat{\chi}_{1}^{2}(t)-s_{1}^{2}}\right)\label{eq:b1}
\end{equation}
where $\hat{\chi}_{1}$ is defined in (\ref{eq:n2}) or (\ref{eq:n4}).
\begin{rem}
	\label{rem:Notably,-when-} When $\hat{\chi}_{1}$ is a positive constant,
	$V_{1}$ becomes a standard barrier Lyapunov function \cite{key-20a,key-3a,key-33} such that if $|s_{1}|<\hat{\chi}_{1}$,
	then $V_{1}\rightarrow+\infty$ as $|s_{1}|\rightarrow\hat{\chi}_{1}$.
	The parameter $\hat{\chi}_{1}$ acts as a barrier for the virtual
	control error $s_{1}$. The purpose of adopting the barrier Lyapunov
	function is to constrain $s_{1}$ in the interval $(-\hat{\chi}_{1},\hat{\chi}_{1})$.
	We can see that if $V_{1}$ is bounded, $s_{1}\in(-\hat{\chi}_{1},\hat{\chi}_{1})$.
	In addition, in Section 3.3 we will show the parameter $\hat{\chi}_{1}$
	remains to be a constant.~~\Square{}
\end{rem}

By (\ref{eq:b1}), we have the following result.
\begin{prop}
	\label{prop:1}Suppose $\hat{\chi}_{1}(t)$ is a positive constant
	and $|s_{1}|<\hat{\chi}_{1}$. Then, by using (\ref{eq:n2}) with
	$i=1$, \textup{$\dot{V}_{1}$} can be expressed as
	\begin{align}
	\dot{V}_{1}\leq & -a_{1}V_{1}^{\frac{1+\alpha}{2}}-Q_{1}s_{1}^{1+\alpha}-\frac{h_{1}(x_1)K_{1}\hat{\Theta}_{1}s_{1}^{1+\alpha}}{\hat{\chi}_{1}^{2}-s_{1}^{2}}\nonumber \\
	& +\frac{U_{1}(F_{1}(x_{1},\hat{\chi}_{1})-h_{1}(x_1)\hat{\Theta}_{1})s_{1}^{1+\alpha}}{(\hat{\chi}_{1}^{2}-s_{1}^{2})^{2+2\alpha}}+c_{12}s_{2}^{1+\alpha}\label{eq:9-3}
	\end{align}
	where $a_{1},Q_{1},c_{12}$ are positive design parameters\textcolor{black}{,}
	$F_{1}(x_{1},\hat{\chi}_{1})$ is an unknown non-negative function
	which comes from the unstructured uncertainties in (\ref{eq:1-1}). 
\end{prop}
\begin{IEEEproof}
	Under the assumption that $\hat{\chi}_{1}(t)$ is a positive constant,
	differentiating $V_{1}$ with respect to time and using (\ref{eq:1-1})
	and Lemma \ref{lem:A1}, we have
	\begin{align}
	\dot{V}_{1}\leq & \frac{s_{1}(h_{1}x_{2}^{*}+f_{1}(x_{1}))}{\hat{\chi}_{1}^{2}-s_{1}^{2}}+\frac{s_{1}h_{1}(x_{2}-x_{2}^{*})}{\hat{\chi}_{1}^{2}-s_{1}^{2}}\nonumber \\
	\leq & \frac{s_{1}h_{1}x_{2}^{*}+s_{1}^{2}\psi_{1}(x_{1})}{\hat{\chi}_{1}^{2}-s_{1}^{2}}+\frac{s_{1}h_{1}(x_{2}-x_{2}^{*})}{\hat{\chi}_{1}^{2}-s_{1}^{2}}\label{eq:15-1-2}
	\end{align}
	where $\psi_{1}(x_{1})$ is an unknown function.
	
	By the inequalities in Lemmas \ref{lem:Q1}-\ref{lem:Q2} and (\ref{eq:n3}),
	it follows that
	\begin{align}
	\dot{V}_{1}\leq & \frac{s_{1}h_{1}x_{2}^{*}}{\hat{\chi}_{1}^{2}-s_{1}^{2}}+\frac{U_{1}s_{1}^{2}\left(\psi_{1}/U_{1}\right)}{\hat{\chi}_{1}^{2}-s_{1}^{2}}\nonumber \\
	& +\frac{h_{1}^{1+q_{2}}c_{11}s_{1}^{1+q_{2}}}{(\hat{\chi}_{1}^{2}-s_{1}^{2})^{1+q_{2}}}+c_{12}s_{2}^{1+q_{2}}\nonumber \\
	= & \frac{s_{1}h_{1}x_{2}^{*}}{\hat{\chi}_{1}^{2}-s_{1}^{2}}+\frac{U_{1}s_{1}^{1+q_{2}}\overline{F}_{1}(x_{1},\hat{\chi}_{1})}{(\hat{\chi}_{1}^{2}-s_{1}^{2})^{2+2\alpha}}+c_{12}s_{2}^{1+\alpha}\label{eq:6}
	\end{align}
	where $c_{11},c_{12}$ are positive known constants, $\overline{F}_{1}(x_{1},\hat{\chi}_{1})=s_{1}^{1-q_{2}}\psi_{1}(\hat{\chi}_{1}^{2}-s_{1}^{2})^{1+2\alpha}/U_{1}+h_{1}^{1+q_{2}}c_{11}(\hat{\chi}_{1}^{2}-s_{1}^{2})^{1+\alpha}$
	is an unknown function. 
	
	With $i=1$, substituting (\ref{eq:n2}) into (\ref{eq:6}), we have
	\begin{align}
	\dot{V}_{1}\leq & -\frac{K_{1}'s_{1}^{1+\alpha}}{\hat{\chi}_{1}^{2}-s_{1}^{2}}-\frac{h_{1}K_{1}\hat{\Theta}_{1}s_{1}^{1+\alpha}}{\hat{\chi}_{1}^{2}-s_{1}^{2}}\nonumber \\
	& +\frac{U_{1}(F_{1}(x_{1},\hat{\chi}_{1})-h_{1}\hat{\Theta}_{1})s_{1}^{1+\alpha}}{(\hat{\chi}_{1}^{2}-s_{1}^{2})^{2+2\alpha}}+c_{12}s_{2}^{1+\alpha}\label{eq:9-3-1}
	\end{align}
	where $K_{1}'$ is a positive parameter and $F_{1}(x_{1},\hat{\chi}_{1})=\overline{F}_{1}(x_{1},\hat{\chi}_{1})+K_{1}'(\hat{\chi}_{1}^{2}-s_{1}^{2})^{1+\alpha}/U_{1}$. 
	
	Note that $V_{1}\leq\frac{s_{1}^{2}}{2(\hat{\chi}_{1}^{2}-s_{1}^{2})}$
	\cite{key-33} and $\frac{1}{\hat{\chi}_{1}^{2}-s_{1}^{2}}\geq\frac{1}{\hat{\chi}_{1}^{2}}$,
	then there exists a sufficiently large $K_{1}'$ such that $-\frac{K_{1}'s_{1}^{1+\alpha}}{\hat{\chi}_{1}^{2}-s_{1}^{2}}\leq-a_{i}V_{i}^{\frac{1+\alpha}{2}}-Q_{i}s_{i}^{1+\alpha}.$
	Using this for (\ref{eq:9-3-1}), we can complete the proof.
\end{IEEEproof}

\textit{Step i}(\textbf{$2\leq i\leq n$}).\textbf{ }Consider the
following Lyapunov function 
\begin{equation}
V_{i}(\overline{x}_{i},t)=\int_{x_{i}^{*}}^{x_{i}}\frac{\upsilon_{i}^{2-q_{i}}(\tau)}{\hat{\chi}_{i}^{2}(t)-\upsilon_{i}^{2}(\tau)}d\tau\label{eq:b2}
\end{equation}
where $\upsilon_{i}(\tau)=\tau^{1/q_{i}}-x_{i}^{*1/q_{i}}$, $\hat{\chi}_{i}(t)\in[0,+\infty)\rightarrow\mathbb{R}$
is the adaptive parameter in (\ref{eq:n2}) or (\ref{eq:n4}). Meanwhile,
$V_{i}$ has the following properties.
\begin{prop}
	\label{prop:9-1-1}(\textcolor{teal}{\cite{key-35}}) Suppose $\hat{\chi}_{i}(t)$
	is a positive constant and $|s_{i}|<\hat{\chi}_{i}$. Then, $V_{i}$
	has the following properties:
	
	1)
	\begin{equation}
	0\leq\frac{2^{1-1/q_{i}}}{\hat{\chi}_{i}^{2}}(x_{i}-x_{i}^{*})^{2/q_{i}}\leq V_{i}\leq\frac{2\hat{\chi}_{i}^{2}}{\hat{\chi}_{i}^{2}-s_{i}^{2}};\label{eq:cond2-1-1-2-1}
	\end{equation}
	
	2) If $x_{i}^{*}$ is bounded, then as $|s_{i}|\rightarrow\hat{\chi}_{i}$,
	$V_{i}\rightarrow+\infty$.
\end{prop}
\begin{rem}
	According to the above result, it can be seen that when $\hat{\chi}_{i}(t)$
	is a constant, the Lyapunov function in (\ref{eq:b2}) is also a barrier
	Lyapunov function like (\ref{eq:b1}). The barrier $\hat{\chi}_{i}$
	is used to constrain the virtual control error $s_{i}$. 
\end{rem}
Next, similar to (\ref{eq:9-3}), we have the following result for
$\dot{V}_{i}$. 
\begin{prop}
	\label{prop:2-1}Suppose $\overline{\hat{\Theta}}_{i}(t),\overline{\hat{\chi}}_{i}(t)$
	are both constant vectors and $|s_{i}|<\hat{\chi}_{i}$ where $\overline{\hat{\Theta}}_{i}\triangleq(\hat{\Theta}_{1},\hat{\Theta}_{2},...,\hat{\Theta}_{i})^{\mathrm{T}}$,
	$\overline{\hat{\chi}}_{i}\triangleq(\hat{\chi}_{1},\hat{\chi}_{2},...,\hat{\chi}_{i})^{\mathrm{T}}$.
	Then, by using (\ref{eq:n2}) or (\ref{eq:n4}), \textup{$\dot{V}_{i}$}
	can be expressed as 
	\begin{align}
	\dot{V}_{i}\leq & -a_{i}V_{i}^{\frac{1+\alpha}{2}}-Q_{i}s_{i}^{1+\alpha}-\frac{h_{i}(\overline{x}_i)K_{i}\hat{\Theta}_{i}s_{i}^{1+\alpha}}{\hat{\chi}_{i}^{2}-s_{i}^{2}}\nonumber \\
	& +\frac{U_{i}s_{i}^{1+\alpha}(F_{i}(\overline{x}_{i},\overline{\hat{\Theta}}_{i-1},\overline{\hat{\chi}}_{i})-h_{i}(\overline{x}_i)\hat{\Theta}_{i})}{(\hat{\chi}_{i}^{2}-s_{i}^{2})^{2+2\alpha}}\nonumber \\
	& +\sum_{j=1}^{i-1}c_{ij}s_{j}^{1+\alpha}+c_{i,i+1}s_{i+1}^{1+\alpha}\label{eq:16-2}
	\end{align}
	where $a_{i},Q_{i},c_{ij}(j=1,2,...,i-1,i+1)$ are positive design
	parameters, \textcolor{black}{$s_{n+1}\triangleq0$.} $F_{i}(\overline{x}_{i},\overline{\hat{\Theta}}_{i-1},\overline{\hat{\chi}}_{i})$
	is an unknown non-negative function which comes from the unstructured
	uncertainties in (\ref{eq:1-1}). 
\end{prop}
\begin{IEEEproof}
	First, under the assumption that $\overline{\hat{\Theta}}_{i}(t),\overline{\hat{\chi}}_{i}(t)$
	are both constant vectors, by resorting to \cite{key-35}, we have
	\begin{align}
	\dot{V}_{i}\leq & \frac{s_{i}^{2-q_{i}}h_{i}x_{i+1}^{*}}{\hat{\chi}_{i}^{2}-s_{i}^{2}}+\frac{U_{i}s_{i}^{1+\alpha}\overline{F}_{i}(\overline{x}_{i},\overline{\hat{\Theta}}_{i-1},\overline{\hat{\chi}}_{i})}{(\hat{\chi}_{i}^{2}-s_{i}^{2})^{2+2\alpha}}\nonumber \\
	& +\sum_{j=1}^{i-1}c_{ij}s_{j}^{1+\alpha}+c_{i,i+1}s_{i+1}^{1+\alpha}\label{eq:12-3-2}
	\end{align}
	where $\overline{F}_{i}(\overline{x}_{i},\overline{\hat{\Theta}}_{i-1},\overline{\hat{\chi}}_{i})$
	is an unknown function. Then, substituting (\ref{eq:n2}) or (\ref{eq:n4})
	into (\ref{eq:12-3-2}), we get 
	\begin{align}
	\dot{V}_{i}\leq & -\frac{K_{i}'s_{i}^{1+\alpha}}{\hat{\chi}_{i}^{2}-s_{i}^{2}}-\frac{h_{i}K_{i}\hat{\Theta}_{i}s_{i}^{1+\alpha}}{\hat{\chi}_{i}^{2}-s_{i}^{2}}\nonumber \\
	& +\frac{U_{i}s_{i}^{1+\alpha}(F_{i}(\overline{x}_{i},\overline{\hat{\Theta}}_{i-1},\overline{\hat{\chi}}_{i})-h_{i}\hat{\Theta}_{i})}{(\hat{\chi}_{i}^{2}-s_{i}^{2})^{2+2\alpha}}\nonumber \\
	& +\sum_{j=1}^{i-1}c_{ij}s_{j}^{1+\alpha}+c_{i,i+1}s_{i+1}^{1+\alpha}\label{eq:16-2-1}
	\end{align}
	where $K_{i}'$ is a positive design parameter. By (\ref{eq:cond2-1-1-2-1})
	and $\frac{1}{\hat{\chi}_{i}^{2}-s_{i}^{2}}\geq\frac{1}{\hat{\chi}_{i}^{2}}$,
	there exists a sufficiently large $K_{i}'$ such that $-\frac{K_{i}'s_{i}^{1+\alpha}}{\hat{\chi}_{i}^{2}-s_{i}^{2}}\leq-a_{i}V_{i}^{\frac{1+\alpha}{2}}-Q_{i}s_{i}^{1+\alpha}.$
	Using this for the above inequality, we can complete the proof.
\end{IEEEproof}
\begin{rem}
	\textcolor{black}{\label{rem:Note-that-}}\textit{\textcolor{black}{(Controller
			design idea)}} Note that $V_{i}(i=1,2,...,n)$ are all barrier Lyapunov
	functions. By using these functions, we expect to constrain all the
	virtual control errors $s_{i}$, states and adaptive parameters in
	a compact set. Thus, there exist unknown positive constants $\overline{F}_{i},\underline{h}_{i}$
	such that $F_{i}(\overline{x}_{i},\overline{\hat{\Theta}}_{i-1},\overline{\hat{\chi}}_{i})\leq\overline{F}_{i}$
	and $|h_{i}(\overline{x}_i)|\geq\underline{h}_{i}>0$. Then, (\ref{eq:16-2}) can
	be written as 
	\begin{align*}
	\dot{V}_{i}\leq & -a_{i}V_{i}^{\frac{1+\alpha}{2}}-Q_{i}s_{i}^{1+\alpha}-\frac{h_{i}(\overline{x}_i)K_{i}\hat{\Theta}_{i}s_{i}^{1+\alpha}}{\hat{\chi}_{i}^{2}-s_{i}^{2}}\\
	& +\frac{U_{i}s_{i}^{1+\alpha}(\overline{F}_{i}-h_{i}(\overline{x}_i)\hat{\Theta}_{i})}{(\hat{\chi}_{i}^{2}-s_{i}^{2})^{2+2\alpha}}+\sum_{j=1}^{i-1}c_{ij}s_{j}^{1+\alpha}+c_{i,i+1}s_{i+1}^{1+\alpha}
	\end{align*}
	The unknown functions/unstructured
	uncertainties $F_{i}(\overline{x}_{i},\overline{\hat{\Theta}}_{i-1},\overline{\hat{\chi}}_{i})$
	become an unknown parameter $\overline{F}_{i}$. Then, according to
	(\ref{eq:8-1}), there exists a sufficiently large $\sigma_{i}$ such
	that $-h_{i}(\overline{x}_i)K_{i}\hat{\Theta}_{i}<0$ and $\overline{F}_{i}-h_{i}(\overline{x}_i)\hat{\Theta}_{i}<0$.
	Hence, the uncertainties can be canceled (see Section 3.3 for details about how the constant bounds for the unknown functions are obtained).
	In addition, since the unknown complicated function $F_{i}(\overline{x}_{i},\overline{\hat{\Theta}}_{i-1},\overline{\hat{\chi}}_{i})$
	is replaced by an unknown parameter $\overline{F}_{i}$, the complexity
	of the controller is reduced considerably. ~~\Square{}
\end{rem}
\begin{rem}
	\textit{\textcolor{black}{\label{rem:(Switching-barrier-Lyapunov}(Switching
			barrier Lyapunov function)}}\textcolor{black}{{} }From the above remark,
	we can see that the key for the controller design is to constrain
	all the virtual control errors and states. However, our case is much
	more difficult than the existing methods \cite{key-20a,key-3a,key-33}, where only \textit{continuous} dynamics are considered.
	In these references, the state trajectories, controller and adaptive
	law are all continuous with respect to time. Therefore, by using barrier
	Lyapunov functions with constant barriers, it is not hard to constrain
	the states. Yet, in our case, the closed loop nonlinear system is
	a \textit{hybrid} system which contains logic-based switching (discrete
	dynamics). Thus, the existing barrier Lyapunov methods will not be
	applicable. In fact, very few works have considered the barrier Lyapunov
	method for hybrid systems.
	
	\textcolor{black}{Specifically, by (\ref{eq:n3}) the virtual control
		error $s_{i}$ is expressed as $s_{i}=x_{i}^{1/q_{i}}-x_{i}^{*1/q_{i}}$.
		Since $x_{i}^{*}$ contains $\hat{\Theta}_{i-1}(t)$ by (\ref{eq:n2}),
		it is discontinuous with respect to time. This implies that $s_{i}$
		is also discontinuous with respect to time. }\textit{\textcolor{black}{Therefore,
			it is possible that at some time instants, $s_{i}$ will jump outside
			the barrier $\hat{\chi}_{i}$ if $\hat{\chi}_{i}$ remains to be a
			constant (see Fig \ref{fig:5} in the simulation example).}}\textcolor{black}{{}
		This implies that the traditional barrier Lyapunov method is not applicable. }
	\textcolor{black}{Therefore, we propose the new Lyapunov functions
		defined in (\ref{eq:b2}). The novelty here is that the barriers $\hat{\chi}_{i}$
		are regarded as adaptive parameters and will be tuned by the logic-based
		switching rule (see Algorithm 1-4)-b) in Section 3.2).} \textcolor{black}{We refer to this kind of Lyapunov
		functions as }\textit{\textcolor{black}{switching barrier Lyapunov
			functions}}\textcolor{black}{. The logic-based switching rule will
		guarantee that the barrier $\hat{\chi}_{i}$ is always larger than
		the virtual control error $s_{i}$.
		This is an essential difference with the existing barrier Lyapunov
		methods \cite{key-20a,key-3a,key-33}, where the virtual
		control error $s_{i}$ is continuous and the barrier is a constant.}
	
	\textcolor{black}{In Section 3.3, we will  prove the switching barrier
		$\hat{\chi}_{i}$ is bounded and  the virtual control errors are constrained, \emph{i.e.,} $|s_i|<\hat{\chi}_{i}(i=1,2,...,n)$. ~~\Square{}}
	%
	%
\end{rem}

\subsection{Logic-based switching rule }

We will present the algorithm for tuning adaptive parameters $\hat{\Theta}_{i}, \hat{\chi}_{i}$
for $i\in\{1,2,...,n\}$. First, for $i\in\{1,2,...,n\}$,
define the following new supervisory functions $\mathcal{S}_{i}(t)$.
\begin{equation}
\mathcal{S}_{i}(t)=V_{i}(\overline{x}_{i},t)-\eta_{i}(t),\label{eq:17-2}
\end{equation}
\begin{equation}
\dot{\eta}_{i}=-a_{i}\eta_{i}^{\frac{1+\alpha}{2}}-Q_{i}s_{i}^{1+\alpha}+\sum_{j=1}^{i-1}c_{ij}s_{j}^{1+\alpha}+c_{i,i+1}s_{i+1}^{1+\alpha}\label{eq:17-1}
\end{equation}
where $V_{i}$ is given by (\ref{eq:b1}) and (\ref{eq:b2}), $s_{i}$
is from (\ref{eq:n1}) or (\ref{eq:n3}) with $s_{n+1}\triangleq0$, $\eta_{i}$ is an
auxiliary variable. $c_{ij}(j=1,2,...,i-1,i+1)$ and $a_{i},Q_{i}$
are positive design parameters. 

Based on the above supervisory functions, the logic-based switching
rule is shown in Algorithm 1 in Table I. 

\begin{figure}
	\begin{centering}
		\includegraphics[scale=0.9]{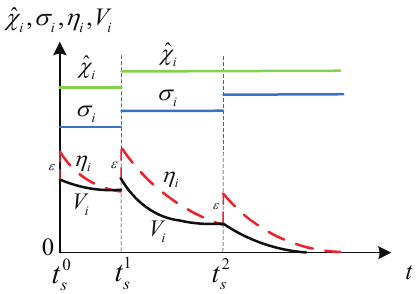}
		\par\end{centering}
	\caption{\label{fig:1-1-1-1}\textcolor{black}{One possible variation of $\hat{\chi}_{i},\sigma_{i},\eta_{i},V_{i}$.
			$\hat{\chi}_{i},\sigma_{i}$ will keep constant when $V_{i}\leq \eta_{i}$.
			As long as the event $V_{i}> \eta_{i}$ occurs, the switching will
			happen and the parameters $\hat{\chi}_{i},\sigma_{i}$ will be updated.
			Then, $\eta_{i}$ is reset to make $V_{i}< \eta_{i}$ and the previous
			procedures will be repeated. (For simplicity, $\hat{\Theta}_i$ and other variables are not shown in this figure.)}}
\end{figure}

Next, we will give some remarks on the algorithm.
\begin{rem}
	\textit{\textcolor{black}{\label{rem:(Idea-of-Algorithm}(Idea of Algorithm
			1)}}\textcolor{black}{{} }The idea of Algorithm 1 is as follows. At each
	time instant $t$, we verify whether or not the supervisory functions
	$\mathcal{S}_{i}(t)>0$. If $\mathcal{S}_{i}(t)\leq0$, the adaptive
	parameters remain the same; otherwise parameters need to be updated.
	The update is conducted in the following way. First, $\sigma_{i},\hat{\Theta}_{i}$
	are updated. The switching signal $\sigma_{i}$ is increased by one
	and the adaptive parameter $\hat{\Theta}_{i}$ is updated by (\ref{eq:8-1}).
	Second, $\hat{\chi}_{i}$ needs to be updated. This is because as
	long as $\hat{\Theta}_{i}$ is updated, the virtual control error
	$s_{i}$ will change accordingly. This may make $s_{i}$ jump at the
	switching time. Therefore, the barrier $\hat{\chi}_{i}$ also needs
	to be updated to guarantee that $\hat{\chi}_{i}$ is always larger
	than $s_{i}$ when $\hat{\Theta}_{i}$ finishes its updating (see
	Remark \ref{rem:(Switching-barrier-Lyapunov}). Third, we reset $\eta_{i}$
	to make sure it is larger than $V_{i}$ after $\hat{\Theta}_{i},\hat{\chi}_{i}$
	have been updated. This will make the supervisory functions $\mathcal{S}_{i}(t)<0$.
	Then, the above procedures will be repeated.
	Fig. \ref{fig:1-1-1-1} shows one possible variations of $\hat{\chi}_{i},\sigma_{i},\eta_{i},V_{i}$. 
	
	Next, we will explain why the adaptive parameters $\hat{\Theta}_{i},\hat{\chi}_{i}$
	are pieceswise constant signals. Note that $\hat{\Theta}_{i},\hat{\chi}_{i}$
	are only updated at the switching time, $i.e.$, the time instant that the
	event $\mathcal{S}_{i}(t)>0$ occurs. When $\mathcal{S}_{i}(t)\leq0$, $\hat{\Theta}_{i},\hat{\chi}_{i}$ will keep constant. Keeping this in mind,
	let $t_{s}^{m}(m=0,1,2,...)$
	denote the proof. At $t_{s}^{m}$,  all the adaptive
	parameters will be updated and $\eta_{i}$ will be reset to make $\mathcal{S}_{i}(t_{s}^{m})<0$
	for $i=1,2,...,n$ (see Algorithm 1-4)). Hence, in the later time
	the adaptive parameters $\hat{\Theta}_{i},\hat{\chi}_{i}$ will keep
	constant until the next event $\mathcal{S}_{i}(t)>0$ occurs. Then,
	parameters will be updated and $\eta_{i}$ will be reset again to
	make $\mathcal{S}_{i}(t)<0$ (see Fig \ref{fig:1-1-1-1}). This indicates
	that $\hat{\Theta}_{i},\hat{\chi}_{i}$ are pieceswise constant signals.
	In addition, since $\mathcal{S}_{i}(t_{s}^{m})<0(i=1,2,...,n)$ after
	the reset of $\eta_{i}$, there exists a small time interval $[t_{s}^{m},t_{s}^{m}+\iota^{m})$
	such that $\mathcal{S}_{i}(t)\leq0$ holds where $\iota^{m}>0$ is
	a small constant. That is the adaptive parameters $\hat{\Theta}_{i},\hat{\chi}_{i}$
	will not change on $[t_{s}^{m},t_{s}^{m}+\iota^{m})$. This also indicates
	that there exists an increasing switching time sequence. See \cite{key-25,key-26,key-4aa} and \cite{key-4-2} for similar idea.
	
	The purpose of Algorithm 1 is to let $\mathcal{S}_{i}(t)=V_{i}(\overline{x}_{i},t)-\eta_{i}(t)\leq0(i=1,...,n)$
	hold forever after a \textit{finite} number of switchings. Then, we have
	$\eta_{i}(t)\geq V_{i}(\overline{x}_{i},t)\geq0(i=1,2,...,n)$ are
	all non-negative (The finite number of switchings and $V_{i} \ge 0$ will be shown in Claim 1b and its proof  in Section 3.3). Then, let $\overline{\eta}\triangleq\sum_{i=1}^{n}\eta_{i}$,
	from (\ref{eq:17-1}) we have
	\begin{align}
	\dot{\overline{\eta}}= & -\sum_{i=1}^{n}a_{i}\eta_{i}^{\frac{1+\alpha}{2}}-\sum_{i=1}^{n}Q_{i}s_{i}^{1+\alpha}\nonumber \\
	& +\sum_{i=1}^{n}\sum_{j=1}^{i-1}c_{ij}s_{j}^{1+\alpha}+\sum_{i=1}^{n}c_{i,i+1}s_{i+1}^{1+\alpha}\nonumber \\
	= & -\sum_{i=1}^{n}a_{i}\eta_{i}^{\frac{1+\alpha}{2}}-\sum_{i=1}^{n}\left(Q_{i}-\sum_{j=i+1}^{n}c_{ji}-c_{i-1,i}\right)s_{i}^{1+\alpha}\label{eq:33-1}
	\end{align}
	where $s_{n+1}\triangleq0$, $\sum_{j=n+1}^{n}c_{jn}\triangleq0$,
	$c_{0i}\triangleq0$. It can be seen that when $Q_{i}$ is sufficiently
	large such that $Q_{i}-\sum_{j=i+1}^{n}c_{ji}-c_{i-1,i}>0$, by Lemma
	\ref{lem:Q2}, we have
	\begin{align}
	\dot{\overline{\eta}}\leq & -\sum_{i=1}^{n}a_{i}\eta_{i}^{\frac{1+\alpha}{2}}\leq-a'\overline{\eta}^{\frac{1+\alpha}{2}}\label{eq:33}
	\end{align}
	where $a'>0$ is a positive constant. This means $\overline{\eta}$
	may converge to zero in finite time. Since $\eta_{i}\geq V_{i}\geq0$
	for $\forall i=1,2,...,n$, this implies that $\eta_{i}$ and $V_{i}$
	may also converge to zero in finite time. By (\ref{eq:n1})-(\ref{eq:n4}), (\ref{eq:b1}), (\ref{eq:b2}) and (\ref{eq:cond2-1-1-2-1}), we can see that all the states will converge to zero in finite time. Please see Claim 1c and its proof in Section 3.3 for details. ~~\Square{}
	\vspace{40pt}
\end{rem}

\begin{algorithm}
	\textit{\underline{Initialization}}
	
	{At $t=0$,}
	\begin{enumerate}[~~1)] 
		\item \textit{Set initial values.} Set design parameters $\varsigma,\varepsilon$
		where $\varsigma>0$ and $\varepsilon>0$ is an arbitrary small constant.
		\textcolor{black}{Set $m:=0$ and the switching time $t_{s}^{m}:=0$.}
		For $i=1,2,...,n$, set $\sigma_{i}(0):=1$ and compute $\hat{\Theta}_{i}(0)$
		by (\ref{eq:8-1}); set $\hat{\chi}_{i}(0)>s_{i}(0)$ and $\eta_{i}(0):=V_{i}(0)+\varepsilon$.
		\item \textit{Output the current} $\sigma_{i},\hat{\Theta}_{i},\hat{\chi}_{i}$. 
	\end{enumerate}
	\textit{\underline{Switching logic}}
	
	\textcolor{black}{At each time instant $t>0$,}
	\begin{enumerate}[~~1)] 
		\item \textit{Compute supervisory functions.} Obtain the current states
		$\overline{x}_{n}(t)$. \textcolor{black}{For $i=1,2,...,n$, set $\sigma_{i}(t):=\sigma_{i}(t^{-})$,
			$\hat{\Theta}_{i}(t):=\hat{\Theta}_{i}(t^{-})$, $\hat{\chi}_{i}(t):=\hat{\chi}_{i}(t^{-})$
			and compute $V_{i},\eta_{i},\mathcal{S}_{i}$ by (\ref{eq:b1}), (\ref{eq:b2}),
			(\ref{eq:17-1}) and (\ref{eq:17-2});}
		\item \textit{Verify parameters update conditions}. For $i=1,2,...,n$,
		check whether or not $\mathcal{S}_{i}(t)>0$ for some $i\in\{1,2,...,n\}$; 
		\item If $\mathcal{S}_{i}(t)\leq0$ for $\forall i=1,2,...,n$, $\hat{\Theta}_{i}$
		and $\hat{\chi}_{i}(i=1,2,...,n)$ are not updated and keep constant.
		\textcolor{black}{Goto 5) to output parameters directly;}
		\item If $\mathcal{S}_{i}(t)>0$ for some $i\in\{1,2,...,n\}$, $\hat{\Theta}_{i}$
		and $\hat{\chi}_{i}(i=1,2,...,n)$ need to be updated.\textcolor{black}{{}
			Set the current time instant to the switching time, $i.e.$, set $m:=m+1,t_{s}^{m}:=t$.
			Then, do the following:}
		\begin{enumerate}[a)] 
			\item \textit{Update} $\hat{\Theta}_{i}$. For $i=1,2,...,n$, if $\mathcal{S}_{i}(t_{s}^{m})>0$,
			set $\sigma_{i}(t_{s}^{m}):=\sigma_{i}(t_{s}^{m})+1$ and update
			$\hat{\Theta}_{i}(t_{s}^{m})$ by (\ref{eq:8-1}); otherwise $\sigma_{i},\hat{\Theta}_{i}$
			are not updated;
			\item \textit{Update} $\hat{\chi}_{i}$. For $i=1,2,...,n$, recompute $s_{i}(t_{s}^{m})$,
			if $|s_{i}(t_{s}^{m})|\geq|\hat{\chi}_{i}(t_{s}^{m})|$, update $\hat{\chi}_{i}(t_{s}^{m}):=|s_{i}(t_{s}^{m})|+\varsigma$
			to make $|s_{i}(t_{s}^{m})|<\hat{\chi}_{i}(t_{s}^{m})$; otherwise
			$\hat{\chi}_{i}$ is not updated;
			\item \textit{Reset} $\eta_{i}$. For $i=1,2,...,n$, recompute $V_{i}(\overline{x}_{i}(t_{s}^{m}),t_{s}^{m})$,
			if $V_{i}(\overline{x}_{i}(t_{s}^{m}),t_{s}^{m})\geq\eta_{i}(t_{s}^{m})$,
			\textcolor{black}{reset $\eta_{i}(t_{s}^{m}):=V_{i}(\overline{x}_{i}(t_{s}^{m}),t_{s}^{m})+\varepsilon$
				to make $\mathcal{S}_{i}(t_{s}^{m})<0$}; otherwise $\eta_{i}$ is
			not updated; 
		\end{enumerate}
		\item \textit{Output the current} $\sigma_{i},\hat{\Theta}_{i},\hat{\chi}_{i}$. 
	\end{enumerate}
	\caption{Logic-based switching rule. }
\end{algorithm}

\begin{rem}
	\textit{\textcolor{black}{\label{rem:(New-supervisory-functions)}(New
			supervisory functions)}} \textcolor{black}{In order to deal with the unstructured uncertainties $h_{i}(\overline{x}_{i}), f_{i}(\overline{x}_{i})$ in each channel of the system (\ref{eq:1-1}). The proposed virtual/real control
		effort $x_{i+1}^{*}$ in (\ref{eq:n2}) or (\ref{eq:n4}) contains adaptive parameters
		$\hat{\chi}_{i},\hat{\Theta}_{i}$ for $i=1,2,...,n$. \textit{Therefore,
			to guarantee the stability, we need to successively show the boundedness
			of the parameters $\hat{\chi}_{i},\hat{\Theta}_{i}$ and state $x_{i}$
			for $i=1,2,...,n$.} This makes the logic-based switching rule in
		the existing works invalid, where the adaptive parameters may only
		exist in the last control effort $x_{n+1}^{*}=u$ \cite{key-25}.} Hence, we propose the new supervisory functions (\ref{eq:17-2})-(\ref{eq:17-1})
	to guide the logic-based switching. The new supervisory functions
	have the following two major differences from the existing methods,
	which are important for the boundedness of the adaptive parameters:
	
	1) In the existing works \cite{key-25,key-4a}, the Lyapunov function is compared with a pre-specified
	time-varying function. However, for (\ref{eq:17-2}), the Lyapunov
	function $V_{i}$ is compared with a dynamic variable $\eta_{i}$
	which is determined by the constructed auxiliary system (\ref{eq:17-1}).
	It relies on the current state information.
	
	2) In \cite{key-25}\textcolor{teal}, only one single supervisory function is used to guide
	the switching for the adaptive parameters. Yet, in our case, we have
	used $n$ different supervisory functions $\mathcal{S}_{i}(t)$ to
	guide the switching for the adaptive parameters in every virtual control
	effort $x_{i+1}^{*}$. 
	
	\textcolor{black}{Note that the new supervisory functions and switching
		barrier Lyapunov function explained in Remark \ref{rem:(Switching-barrier-Lyapunov}
		make the the proposed method has some substantial differences with
		the existing methods, $e.g.$, \cite{key-25,key-4a}
		and \cite{key-3a}. This is reflected in the stability analysis
		in Section 3.3, where we propose a new 
		3-Claims procedure to show the finite time stability. ~~\Square{}}
\end{rem}

\subsection{Main result and stability analysis}

Based on the analysis in Sections 3.1 and 3.2, we have the following
main result.
\begin{thm}
	\label{thm:Consider-the-nonlinear-1-1}Consider the nonlinear system
	in (\ref{eq:1-1}). Then, the controller (\ref{eq:n1})-(\ref{eq:n4}) with Algorithm
	1 can guarantee that:
	
	1) All the signals in the closed-loop system are bounded for $\forall t\in[0,+\infty)$,
	and; 
	
	2) All the states will converge to zero in finite time.
\end{thm}
The proof for the above result will be presented in this subsection.
According to Remark \ref{rem:(Idea-of-Algorithm}, we can define a
switching time sequence $\{0=t_{s}^{0}<t_{s}^{1}<...<t_{s}^{m}<...\le +\infty\}$
such that
\begin{equation}
t_{s}^{m+1}=\inf\{t|t\geq t_{s}^{m},\mathcal{S}_{i}(t)>0,i\in\{1,2,...,n\}\}.\label{eq:20-4}
\end{equation}
During time interval $[t_{s}^{m},t_{s}^{m+1})$, the supervisory function
satisfies $\mathcal{S}_{i}(t)=V_{i}(\overline{x}_{i},t)-\eta_{i}(t)\leq0$
for $\forall i=1,2,...,n$. \textcolor{black}{Meanwhile, $\sigma_{i}, \hat{\Theta}_{i},\hat{\chi}_{i}(i=1,2,...,n)$
	are all constants on $[t_{s}^{m},t_{s}^{m+1})$,  \emph{i.e.},  $\sigma_{i}(t)=\sigma_{i}(t_{s}^{m})$, $\hat{\Theta}_{i}(t)=\hat{\Theta}_{i}(t_{s}^{m})$, $\hat{\chi}_{i}(t)=\hat{\chi}_{i}(t_{s}^{m})$ for $\forall t\in[t_{s}^{m},t_{s}^{m+1})$.}

The proof will be obtained by proving the following three claims,
\textit{i.e.}, Claims 1a, 1b and 1c. Claim 1a tries to show the boundedness
of signals in the system if the number of switchings is finite. Next,
Claim 1b attempts to show  the number of switchings is indeed finite. Finally,
Claim 1c proves the finite time stability.\smallskip{}

\noindent \textbf{Claim 1a.} \textit{For any finite integer $m$,
	we have}

\textit{1) The closed loop nonlinear system admits continuous solution
	$\overline{x}_{n}(t)$ on $[t_{s}^{0},t_{s}^{m+1})$; }

\textit{2) There exists a positive constant $\delta_{i}^{m}$ such
	that $|s_{i}(t)|\leq\hat{\chi}_{i}(t_{s}^{m})-\delta_{i}^{m}$ for
	$\forall i=1,2,...,n$ on $[t_{s}^{m},t_{s}^{m+1})$; }

\textit{3) All the signals in the system are bounded on $[t_{s}^{m},t_{s}^{m+1})$. }
\begin{IEEEproof}
	According to Algorithm 1 and Remark \ref{rem:(Idea-of-Algorithm},
	we know during each time interval \textit{$[t_{s}^{m},t_{s}^{m+1})$},
	the initial condition satisfies \textit{$|s_{i}(t_{s}^{m})|<\hat{\chi}_{i}(t_{s}^{m})$}
	and the barrier $\hat{\chi}_{i}$ remains to be constant. Hence, (\ref{eq:b1})
	and (\ref{eq:b2}) become traditional barrier Lyapunov functions on
	each \textit{$[t_{s}^{m},t_{s}^{m+1})$}. Therefore, according to
	the theory of barrier Lyapunov functions \cite{key-20a} and the fact that $V_{i}(\overline{x}_{i},t)\leq\eta_{i}(t)$
	on each \textit{$[t_{s}^{m},t_{s}^{m+1})$}, we can show the virtual
	control error $s_{i}(t)$ is constrained for each\textit{ $[t_{s}^{m},t_{s}^{m+1})$},
	$i.e.$, \textit{$|s_{i}(t)|\leq\hat{\chi}_{i}(t_{s}^{m})-\delta_{i}^{m}(i=1,2,...,n)$}
	for\textit{ $\forall t\in[t_{s}^{m},t_{s}^{m+1})$}. Detail proofs
	are put in Appendix B. 
\end{IEEEproof}
\noindent \textbf{Claim 1b.} \textit{1) The number of switchings is finite;}

\textit{2) The closed loop nonlinear system admits continuous solution
	$\overline{x}_{n}(t)$ on $[0,+\infty)$; }

\textit{3) All the signals in the system are bounded on $[0,+\infty)$. }
\begin{IEEEproof}
	\textcolor{black}{Note that from Remark \ref{rem:(Idea-of-Algorithm} and Claim 1a, we know during each time interval }\textit{\textcolor{black}{$[t_{s}^{m},t_{s}^{m+1})$}}\textcolor{black}{,
		the adaptive paramters keep constant, and }\textcolor{black}{$|s_{i}(t)|<\hat{\chi}_{i}(t_{s}^{m})=\hat{\chi}_{i}(t)(i=1,...,n)$ with  $t \in [t_{s}^{m},t_{s}^{m+1})$.}
	\textcolor{black}{
		Thus, $V_{i}\geq0(i=1,...,n)$ by (\ref{eq:b1})
		and Proposition \ref{prop:9-1-1}. Meanwhile, the propositions in Section 3.1 are all
		valid for each }\textit{\textcolor{black}{$[t_{s}^{m},t_{s}^{m+1})$}}\textcolor{black}{.
		Keeping this in mind, we will first prove the number of switchings is
		finite. The proof is divided into the following steps. These steps
		correspond to the $n$ steps Lyapunov functions analysis in Section
		3.1. }
	
	\textit{Step 1.} We will prove $\hat{\chi}_{1},\hat{\Theta}_{1}$
	have finite  numbers of switchings.
	
	1) Show $\hat{\chi}_{1}(t)$ does not switch on $[t_{s}^{0},t_{s}^{m+1})$. 
	
	According to Claim 1a, we know $|s_{1}(t)|\leq\hat{\chi}_{1}(t_{s}^{m})-\delta_{1}^{m}$
	on $[t_{s}^{m},t_{s}^{m+1})$ for any finite integer $m$, and $s_{1}(t)=x_{1}(t)$
	is continuous on $[t_{s}^{0},t_{s}^{m+1})$. Then, according to Algorithm
	1-4) in Switching logic, $\hat{\chi}_{1}$ will not be updated on
	$[t_{s}^{0},t_{s}^{m+1})$ since $s_{1}(t)$ will never transgress
	the barrier $\hat{\chi}_{1}(t_{s}^{0})$.
	
	2) Show $\hat{\Theta}_{1}(t)$ has a finite  number of switchings. 
	
	This is proved by contradiction. If this is not true, then $\hat{\Theta}_{1}$
	will switch infinite times. 
	
	From Claim 1a and the fact that $\hat{\chi}_{1}(t)=\hat{\chi}_{1}(t_{s}^{0})$ with $t \in [t_{s}^{0},t_{s}^{m+1})$,
	we have $|s_{1}(t)|=|x_{1}(t)|<\hat{\chi}_{1}(t_{s}^{0})$ is bounded
	on $[t_{s}^{0},t_{s}^{m+1})$. It follows that on $[t_{s}^{0},t_{s}^{m+1})$,
	the unstructured uncertainties $h_{1}(x_{1})$ and $F_{1}(x_{1},\hat{\chi}_{1})$
	in (\ref{eq:9-3}) satisfy
	\[
	|h_{1}(x_{1})|\geq\underline{h}_{\ensuremath{1}}>0,
	\]
	\[
	0\leq F_{1}(x_{1},\hat{\chi}_{1})\leq\overline{F}_{1}
	\]
	where $\underline{h}_{\ensuremath{1}},\overline{F}_{1}$ are unknown
	constants irrelevant with the number of switchings $m$. 
	
	Then, from the tuning rule (\ref{eq:8-1}) and the assumption of infinite
	switchings, we can conclude that there exists a sufficiently
	large finite integer $m_{1}$ such that at switching time $t_{s}^{m_{1}}$,
	we have
	\begin{align*}
	\mathrm{sgn}(\hat{\Theta}_{1}(t_{s}^{m_{1}})) & =\mathrm{sgn}(h_{1}(x_{1})),\\
	-h_{1}K_{1}\hat{\Theta}_{1}(t_{s}^{m_{1}}) & \leq -\underline{h}_{1}K_{1}|\hat{\Theta}_{1}(t_{s}^{m_{1}})|<0,\\
	F_{1}(x_{1},\hat{\chi}_{1})-h_{1}\hat{\Theta}_{1}(t_{s}^{m_{1}}) & \leq\overline{F}_{1}-\underline{h}_{1}|\hat{\Theta}_{1}(t_{s}^{m_{1}})|<0
	\end{align*}
	This implies that at switching time $t_{s}^{m_{1}}$, (\ref{eq:9-3})
	will become
	\begin{align}
	\dot{V}_{1}\leq & -a_{1}V_{1}^{\frac{1+\alpha}{2}}-Q_{1}s_{1}^{1+\alpha}+c_{12}s_{2}^{1+\alpha}\label{eq:27-3-2}
	\end{align}
	with $a_{1},Q_{1}$ defined in
	(\ref{eq:17-1}).
	
	On the other hand, the auxiliary variable $\eta_{1}(t)$ in (\ref{eq:17-1})
	satisfies
	\begin{align}
	\dot{\eta}_{1}= & -a_{1}\eta_{1}^{\frac{1+\alpha}{2}}-Q_{1}s_{1}^{1+\alpha}+c_{12}s_{2}^{1+\alpha}\label{eq:27-3-2-1}
	\end{align}
	where $\eta_{1}(t_{s}^{m_{1}})\geq V_{1}(x_{1}(t_{s}^{m_{1}}),t_{s}^{m_{1}})$
	according to Algorithm 1-4)-c) in Switching logic.
	
	From Lemma \ref{lem:C1} and (\ref{eq:27-3-2})-(\ref{eq:27-3-2-1}), we know $V_{1}(x_{1}(t),t)\leq\eta_{1}(t)$
	will hold on $[t_{s}^{m_{1}},t_{s}^{m+1})$ for any $m+1>m_{1}$ without
	resetting $\eta_{1}(t)$. This means that $\hat{\Theta}_{1}$ will
	not be updated after $t_{s}^{m_{1}}$ which contradicts the fact that
	$\hat{\Theta}_{1}$ has an infinite number of switchings. 
	
	\textit{Step 2.} We will prove $\hat{\chi}_{2},\hat{\Theta}_{2}$
	have finite numbers of switchings. 
	
	The proof will be conducted on $[t_{s}^{m_{1}},t_{s}^{m+1})$ where
	$t_{s}^{m_{1}}$ denotes the time instant when $\hat{\chi}_{1},\hat{\Theta}_{1}$
	stop switching and keep constant.
	
	1) Show $\hat{\chi}_{2}(t)$ does not switch on $[t_{s}^{m_{1}},t_{s}^{m+1})$. 
	
	Note that $\hat{\Theta}_{1}$ is not updated on $[t_{s}^{m_{1}},t_{s}^{m+1})$.
	Meanwhile, according to Claim 1a, we know $x_{2}$ is continuous on
	$[t_{s}^{m_{1}},t_{s}^{m+1})$. Then, by (\ref{eq:n2}) and (\ref{eq:n3}),
	$s_{2}$ is continuous on $[t_{s}^{m_{1}},t_{s}^{m+1})$. Also by
	Claim 1a, we have $|s_{2}(t)|\leq\hat{\chi}_{2}(t_{s}^{m})-\delta_{2}^{m}$
	on $[t_{s}^{m},t_{s}^{m+1})$ for any finite integer $m$. Therefore,
	according to Algorithm 1-4) in Switching logic, $\hat{\chi}_{2}$
	will not be updated on $[t_{s}^{m_{1}},t_{s}^{m+1})$ since $s_{2}(t)$
	will never transgress the barrier $\hat{\chi}_{2}(t_{s}^{m_{1}})$. 
	
	2) Show $\hat{\Theta}_{2}(t)$ has a finite  number of switchings. 
	
	This is proved by contradiction. We suppose $\hat{\Theta}_{2}(t)$
	will switch infinite times. 
	
	First, since $\hat{\chi}_{2}$ is not  updated on $[t_{s}^{m_{1}},t_{s}^{m+1})$,
	we have $|s_{2}(t)|<\hat{\chi}_{2}(t_{s}^{m_{1}})$ with $t \in [t_{s}^{m_{1}},t_{s}^{m+1})$ by Claim 1a. In addition,
	from Claim 1a, we know $\hat{\chi}_{2}(t_{s}^{m_{1}})$ is bounded. 
	
	Using (\ref{eq:27-3-2-1}) in Step 1 we know $\eta_{1}$ satisfies
	\begin{align*}
	\dot{\eta}_{1}= & -a_{1}\eta_{1}^{\frac{1+\alpha}{2}}-Q_{1}s_{1}^{1+\alpha}+c_{12}s_{2}^{1+\alpha}\\
	\leq & -a_{1}\eta_{1}^{\frac{1+\alpha}{2}}+c_{12}\hat{\chi}_{2}^{1+\alpha}(t_{s}^{m_{1}})
	\end{align*}
	on $[t_{s}^{m_{1}},t_{s}^{m+1})$. 
	
	Therefore, it can be concluded that $0\leq V_{1}\leq\eta_{1}$ is
	bounded by a constant irrelevant with $m$ (see Corollary 1 in \cite{key-3}). Then, from the barrier Lyapunov function (\ref{eq:b1}),
	we know $|s_{1}(t)|=|x_{1}(t)|\leq\hat{\chi}_{1}(t_{s}^{0})-\delta_{1}$
	with a positive constant $\delta_{1}$ irrelevant with $m$. Also
	from (\ref{eq:n2}) and (\ref{eq:n3}), we know $x_{2}^{*},x_{2}$
	are both bounded by constants irrelevant with $m$.
	
	Hence, we conclude that on $[t_{s}^{m_{1}},t_{s}^{m+1})$, $h_{2}(\overline{x}_{2})$
	and $F_{2}(\overline{x}_{2},\hat{\Theta}_{1},\hat{\chi}_{1})$ in
	(\ref{eq:16-2}) satisfy
	\[
	|h_{2}(\overline{x}_{2})|\geq\underline{h}_{\ensuremath{2}}>0,
	\]
	\[
	0\leq F_{2}(\overline{x}_{2},\hat{\Theta}_{1},\overline{\hat{\chi}}_{2})\leq\overline{F}_{2}
	\]
	where $\underline{h}_{\ensuremath{2}},\overline{F}_{2}$ are positive
	constants irrelevant with $m$. Here, we also use the fact that $\hat{\chi}_{1},\hat{\chi}_{2},\hat{\Theta}_{1}$
	are constants on $[t_{s}^{m_{1}},t_{s}^{m+1})$.
	
	Then, from tuning rule (\ref{eq:8-1}), there exists a finite integer
	$m_{2}\geq m_{1}$ such that at switching time $t_{s}^{m_{2}}$, we
	have
	\begin{align*}
	\mathrm{sgn}(\hat{\Theta}_{2}(t_{s}^{m_{2}})) & =\mathrm{sgn}(h_{2}(\overline{x}_{2})),\\
	-h_{2}K_{2}\hat{\Theta}_{2}(t_{s}^{m_{2}}) & \leq -\underline{h}_{2}K_{2}|\hat{\Theta}_{2}(t_{s}^{m_{2}})|<0,\\
	F_{2}(\overline{x}_{2},\hat{\Theta}_{1},\overline{\hat{\chi}}_{2})-h_{2}\hat{\Theta}_{2}(t_{s}^{m_{2}}) & \leq\overline{F}_{2}-\underline{h}_{2}|\hat{\Theta}_{2}(t_{s}^{m_{2}})|<0
	\end{align*}
	This implies that (\ref{eq:16-2}) will become
	\begin{align}
	\dot{V}_{2}\leq & -a_{2}V_{2}^{\frac{1+\alpha}{2}}-Q_{2}s_{2}^{1+\alpha}+c_{21}s_{1}^{1+\alpha}+c_{23}s_{3}^{1+\alpha}\label{eq:44}
	\end{align}
	at switching time $t_{s}^{m_{2}}$ with $a_{2},Q_{2}$ defined in
	(\ref{eq:17-1}).
	
	On the other hand, the auxiliary variable $\eta_{2}(t)$ in (\ref{eq:17-1})
	satisfies
	\begin{align}
	\dot{\eta}_{2}= & -a_{2}\eta_{2}^{\frac{1+\alpha}{2}}-Q_{2}s_{2}^{1+\alpha}+c_{21}s_{1}^{1+\alpha}+c_{23}s_{3}^{1+\alpha}\label{eq:44-1-1}
	\end{align}
	where $\eta_{2}(t_{s}^{m_{2}})\geq V(\overline{x}_{2}(t_{s}^{m_{2}}),t_{s}^{m_{2}})$
	according to Algorithm 1-5) in Switching logic.
	
	From Lemma \ref{lem:C1} and (\ref{eq:44})-(\ref{eq:44-1-1}), we know $V_{2}(\overline{x}_{2}(t),t)\leq\eta_{2}(t)$
	will hold on $[t_{s}^{m_{2}},t_{s}^{m+1})$ for any $m+1>m_{2}$ without
	resetting. This means that $\hat{\Theta}_{2}$ will not be updated
	after $t_{s}^{m_{2}}$ which contradicts the fact that $\hat{\Theta}_{2}$
	has an infinite number of proof. 
	
	\textit{Step i}(\textbf{$3\leq i\leq n$}). By repeating the above
	procedures, we can show all the parameters $\hat{\Theta}_{i},\hat{\chi}_{i}(i=1,2,...,n)$
	have  finite numbers of switchings. Statement 1) in Claim 1b is proved
	
	Next, for Statements 2)-3) in Claim 1b, according to Claim 1a, they
	hold naturally when the number of switchings is finite. In fact, there must exist a finite integer $m_{n}$ such that the switching time  $t_{s}^{m_{n}+1} = +\infty$. The proof is completed.
\end{IEEEproof}
\noindent \textbf{Claim 1c.} \textit{All the states will converge
	to zero in finite time.}
\begin{IEEEproof}
	\textcolor{black}{From Claim 1a, we know $|s_{i}(t)|<\hat{\chi}_{i}(t_{s}^{m})=\hat{\chi}_{i}(t)(i=1,...,n)$
		for $\forall t \in [t_{s}^{m},t_{s}^{m+1})$. Then, we have $V_{i}\geq0(i=1,...,n)$
		by (\ref{eq:b1}),
		(\ref{eq:b2}) and Proposition \ref{prop:9-1-1}. From Claim 1b, we know the number of switchings is finite. This implies that there exists a finite  switching time 
		$t_{s}^{m_{n}}$ where  $m_n$ is a finite integer and $t_{s}^{m_{n}+1}=+\infty$. When $t\in [t_{s}^{m_{n}}, t_{s}^{m_{n}+1})=[t_{s}^{m_{n}},+\infty)$,  $0\leq V_{i}\leq\eta_{i}$
		holds for $\forall i\in\{1,2,...,n\}$ without resetting $\eta_{i}$.}
	Then, on $[t_{s}^{m_{n}},+\infty)$, by (\ref{eq:17-1})-(\ref{eq:33})
	we have
	\[
	\dot{\overline{\eta}}\leq-a'\overline{\eta}^{\gamma}
	\]
	where $a'>0$ is a positive parameter, $\gamma\triangleq\frac{1+\alpha}{2}\in(0,1)$.
	
	Therefore, during time interval $[t_{s}^{m_{n}},+\infty)$, we have\textcolor{black}{
		\begin{align*}
		{\color{black}{\color{black}0\leq}} & {\color{black}{\color{black}\sum_{i=1}^{n}V_{i}\leq\overline{\eta}}}(t)\\
		\leq & [\overline{\eta}^{1-\gamma}(t_{s}^{m_{n}})-a'(1-\gamma)(t-t_{s}^{m_{n}})]^{\frac{1}{1-\gamma}}.
		\end{align*}
	}Note that by Claim 1b, all the signals including $\overline{\eta}(t_{s}^{m_{n}})$
	are bounded on $[0,+\infty)$. Hence, we can conclude that after finite
	time $t_{s}^{m_{n}}+\frac{\eta^{1-\gamma}(t_{s}^{m_{n}})}{a'(1-\gamma)}$,
	$\overline{\eta}$ and $V_{i}(i=1,2,...,n)$ will converge to zero. By (\ref{eq:n1})-(\ref{eq:n4}), (\ref{eq:b1}), (\ref{eq:b2}) and (\ref{eq:cond2-1-1-2-1}), we know 
	$s_1=x_1, x_{2}^{*}, x_2, s_2, x_{3}^{*}, x_3, ..., x_n$  will also converge to zero in finite time.
	The proof is completed.
\end{IEEEproof}

\section{Examples}
\begin{figure}
	\begin{centering}
		\includegraphics[scale=0.37]{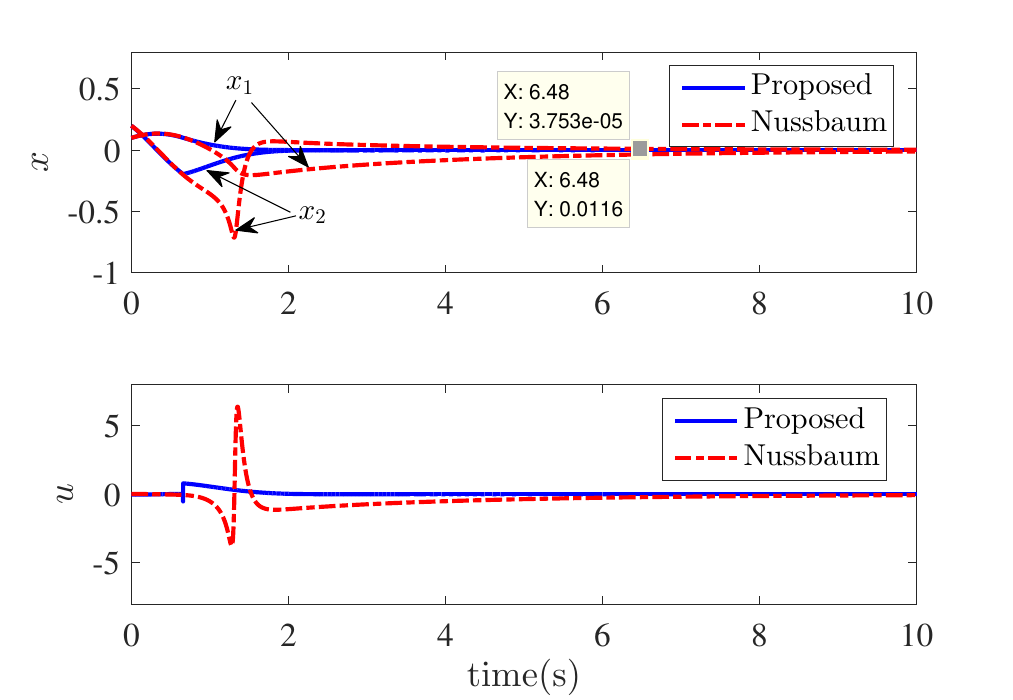}
		\par\end{centering}
	\caption{\label{fig:6}Control performance comparison for Case A.}
\end{figure}

\begin{figure}
	\begin{centering}
		\includegraphics[scale=0.37]{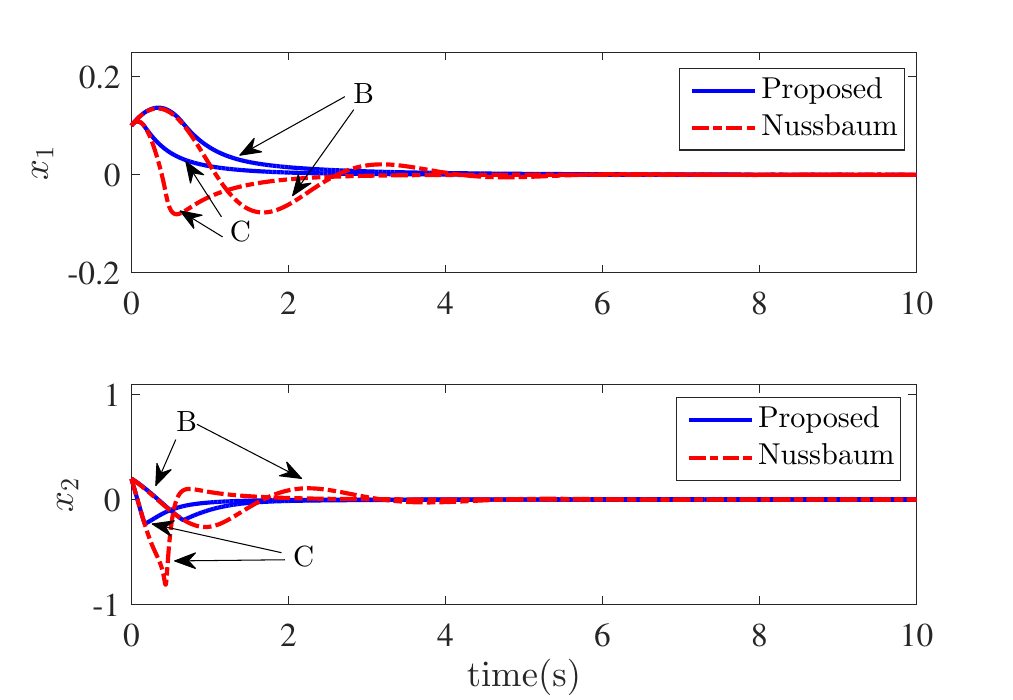}
		\par\end{centering}
	\caption{\label{fig:7}Control performance comparison for Cases B and C.}
\end{figure}

\begin{figure}
	\begin{centering}
		\includegraphics[scale=0.37]{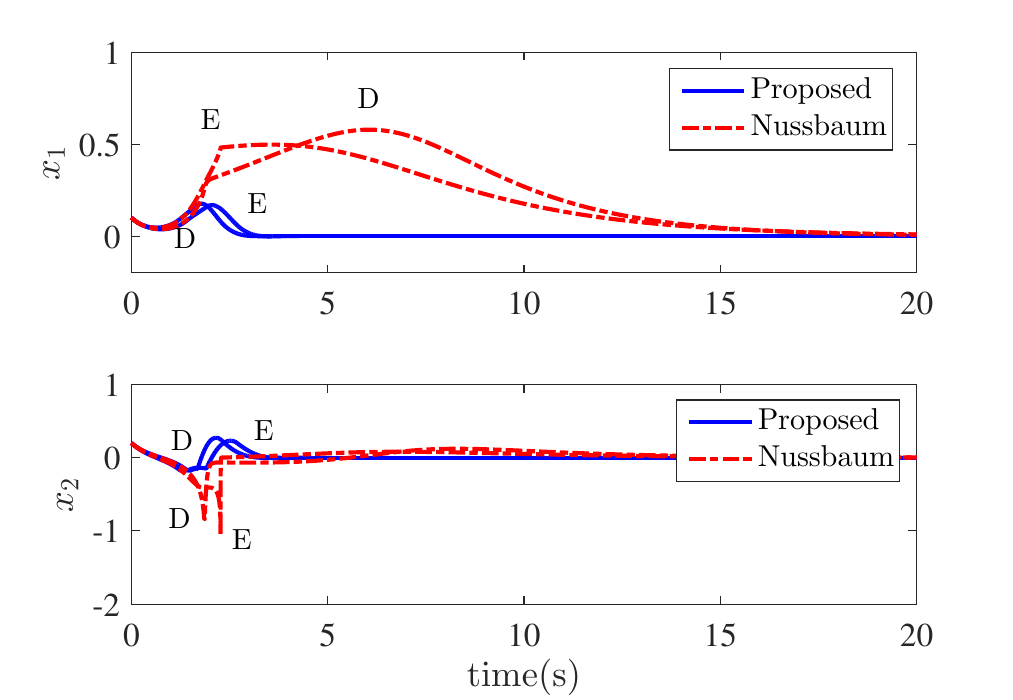}
		\par\end{centering}
	\caption{\label{fig:8}Control performance comparison for Cases D and E.}
\end{figure}

\begin{figure}
	\begin{centering}
		\includegraphics[scale=0.35]{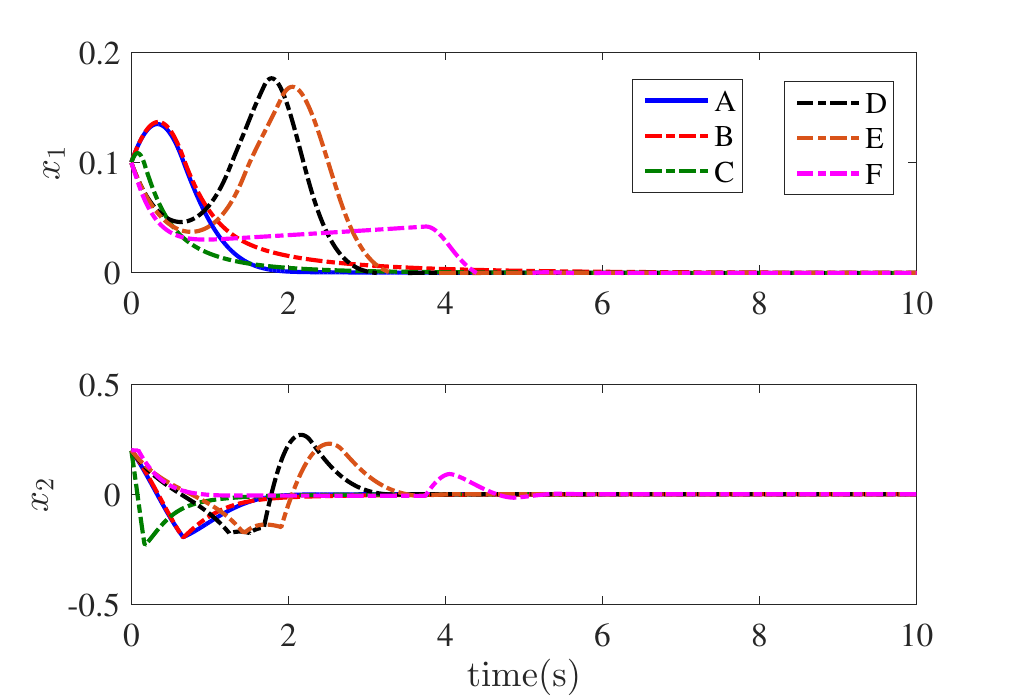}
		\par\end{centering}
	\caption{\label{fig:3}Control performance for second order system.}
\end{figure}

\begin{figure}
	\begin{centering}
		\includegraphics[scale=0.35]{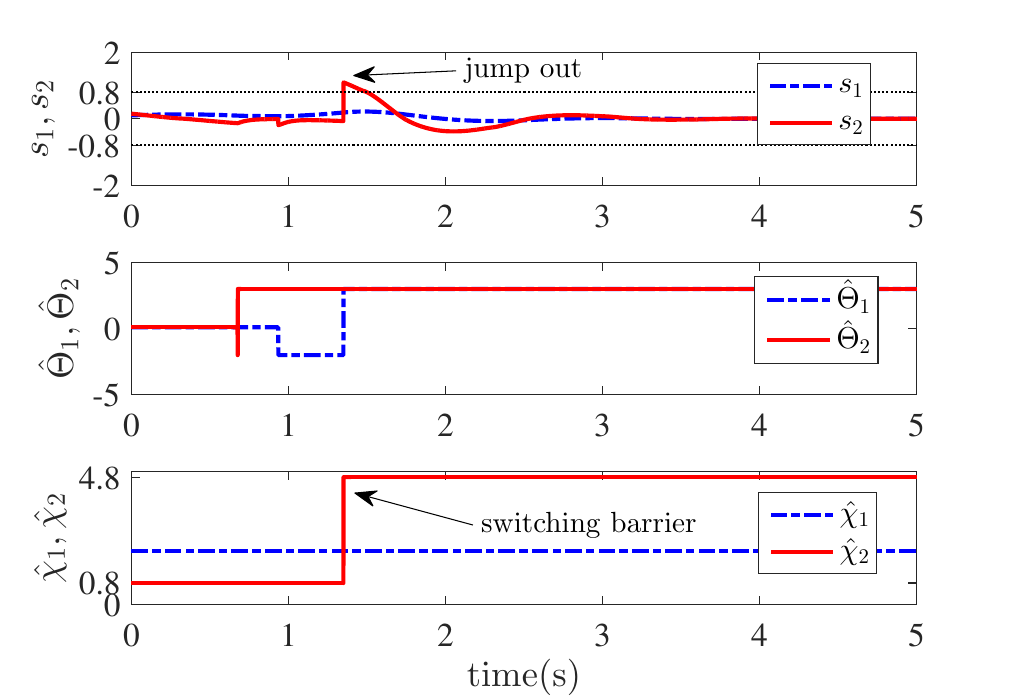}
		\par\end{centering}
	\caption{\label{fig:5}\textcolor{black}{Variations of $s_{i},\hat{\Theta}_{i},\hat{\chi}_{i}(i=1,2)$.
			It can be seen that $|s_{2}|$ is constrained in the barrier $\hat{\chi}_{2}(t)=0.8$
			during time interval $[0,1.35)$ and jump outside the barrier $\hat{\chi}_{2}(t)$
			at time instant $1.35s$ due to $\hat{\Theta}_{1}$ is updated. Then,
			by the proposed method, $\hat{\chi}_{2}$ will switch to $4.8$ to
			contain $s_{2}$ in the later time to guarantee the finite time stability. }}
\end{figure}

\textcolor{black}{In this section, an illustrative example is presented. Note that some further discussions
	about convergence speed, control overshoot, controller parameters
	selection, comparison with existing asymptotic control methods and control of
	third order nonlinear systems are put in Appendix F in the supplementary file.}
\begin{example}
	\label{exa:Consider-a-robot}Given a second order nonlinear system
	by (\ref{eq:1-1}) with $n=2$, we consider the following six cases:
	\begin{align*}
	\mathrm{Case}\thinspace\mathrm{A}: & h_{1}=1,h_{2}=0.8,f_{1}=\overline{f}_{1},f_{2}=\overline{f}_{2};\\
	\mathrm{Case}\thinspace\mathrm{B}: & h_{1}=1,h_{2}=-0.8,f_{1}=\overline{f}_{1},f_{2}=\overline{f}_{2};\\
	\mathrm{Case}\thinspace\mathrm{C}: & h_{1}=1,h_{2}=0.8,f_{1}=5\overline{f}_{1},f_{2}=5\overline{f}_{2};\\
	\mathrm{Case}\thinspace\mathrm{D}: & h_{1}=-1,h_{2}=0.8,f_{1}=\overline{f}_{1},f_{2}=\overline{f}_{2};\\
	\mathrm{Case}\thinspace\mathrm{E}: & h_{1}=-1,h_{2}=-0.8,f_{1}=\overline{f}_{1},f_{2}=\overline{f}_{2};\\
	\mathrm{Case}\thinspace\mathrm{F}: & h_{1}=-1,h_{2}=0.8,f_{1}=f_{1}^{*},f_{2}=f_{2}^{*}
	\end{align*}
	where $\overline{f}_{1}\triangleq0.1\sin(x_{1})x_{1}$; $\overline{f}_{2}\triangleq-4.9\sin(x_{1})+0.05\sin(x_{1})e^{-x_{2}}+0.1\sin(x_{2})x_{2}^{2}.$
	$f_{1}^{*}\triangleq-3x_{1}^{2}/2-x_{1}^{3}/2$ and $f_{2}^{*}\triangleq0.1\sin(x_{2})$.
	Case A is the nominal case. It can be used to describe the dynamics
	of a single link robot manipulator \cite{key-23a}. Other five situations represent the variation
	of control directions and modeling uncertainties. Specifically, Case
	F indicates that the whole system has been changed into another form.
	The initial conditions are $x_{1}(0)=0.1,x_{2}(0)=0.2.$ For the controller
	design, we assume $h_{1},h_{2},f_{1},f_{2}$ are all unknown. 
	

	\textit{1) Effectiveness of the logic-based switching}
	
	The controller is designed by  (\ref{eq:n1})-(\ref{eq:n4}). 
	The controller parameters are set as: $K_{1}=U_{1}=K_{2}=U_{2}=1$,
	$\alpha=41/49$, $\hat{\chi}_{1}=\hat{\chi}_{2}(0)=2$. $\theta_{1}(0)=\theta_{2}(0)=0.1$
	and for $\sigma_{i}(t)\geq1(i=1,2)$, we have $\theta_{i}(\sigma_{i})=(-1)^{\sigma_{i}(t)}(\iota_{i1}+\iota_{i2}\sigma_{i})$
	where $\iota_{i1}=\iota_{i2}=1.$ $\hat{\Theta}_{1},\hat{\Theta}_{2}\text{,\ensuremath{\hat{\chi}_{1}},\ensuremath{\hat{\chi}_{2}} }$
	are updated by Algorithm 1 with $\varsigma=4,\varepsilon=0.01$, $a_{1}=a_{2}=0.2,$
	$Q_{1}=Q_{2}=c_{11}=c_{21}=1$. The control performance is shown in Fig. \ref{fig:3}. It can be seen
	that both states converge to zero in a very short time for the above
	six situations. This implies that the finite time stability has been
	achieved despite multiple unknown control directions and unstructured
	uncertainties.
	
	Next, set $\hat{\chi}_{2}(0)=0.8,\hat{\chi}_{1}(0)=2$. The variations
	of $s_{i},\hat{\Theta}_{i},\hat{\chi}_{i}(i=1,2)$ are shown in Fig.
	\ref{fig:5}. Note that $s_{1}$ is constrained into the tube $[-2,2]$.
	This shows the validity of the barrier Lyapunov function $V_{1}$
	in (\ref{eq:b1}). Also we can see $s_{2}$ is constrained into the
	tube $[-0.8,0.8]$ during time interval $[0,1.35)$ and is larger
	than $\hat{\chi}_{2}(0)=0.8$ at time instant $1.35s$. Therefore,
	$\hat{\chi}_{2}$ jumps to $4.8$ to contain $s_{2}$ in the later
	time. The reason for $s_{2}$ jumping outside $0.8$ is that $\hat{\Theta}_{1}$
	is updated at $1.35s$. Note that though $s_{2}$ transgresses over
	the barrier $\hat{\chi}_{2}(0)=0.8$, $s_{1},s_{2}$ still can converge
	to zero in finite time. All these show the effectiveness of the switching
	barrier Lyapunov function.

	\textit{2) Comparision with Nussbaum-gain method}
	
Consider the system described by Example \ref{exa:Consider-a-robot}.
Given a finite time controller by (\ref{eq:n1})-(\ref{eq:n4}) with the same parameters in Example 1,
and another controller designed by Nussbaum-gain technique:
\begin{align*}
u & =\mathcal{N}_{2}(\xi_{2})v_{2},\\
v_{2} & =K_{2}s_{2}+\frac{U_{2}s_{2}}{\hat{\chi}_{2}^{2}-s_{2}^{2}},\\
\dot{\xi}_{2} & =\frac{s_{2}v_{2}}{\hat{\chi}_{2}^{2}-s_{2}^{2}}
\end{align*}
where
\begin{align*}
s_{2} & =x_{2}-x_{2}^{*},\\
x_{2}^{*} & =\mathcal{N}_{1}(\xi_{1})v_{1},\\
v_{1} & =K_{1}s_{1}+\frac{U_{1}s_{1}}{\hat{\chi}_{1}^{2}-s_{1}^{2}},\\
\dot{\xi}_{1} & =\frac{s_{1}v_{1}}{\hat{\chi}_{1}^{2}-s_{1}^{2}}.
\end{align*}
$\mathcal{N}_{i}(\xi_{i})(i=1,2)$ are Nussbaum functions such that
$\mathcal{N}_{i}(\xi_{i})=\xi_{i}^{2}\cos(\xi_{i}).$ $K_{1},U_{1},K_{2},U_{2},\hat{\chi}_{1},\hat{\chi}_{2}$
are positive design parameters. The parameters are set as: $K_{1}=U_{1}=1,K_{2}=U_{2}=4,\hat{\chi}_{1}=2,\hat{\chi}_{2}=1$.
Using these parameters, a satisfactory control performance can be
obtained in the nominal case. This controller is a variation of the
method in \cite{key-20}, which adapts to the
unstructured uncertainties. 

Fig. \ref{fig:6} shows the state trajectories and variation of control
effort in Case A. We can see that the finite time control method has
a faster convergence speed and higher precision than Nussbaum-gain
method. In fact, the states by finite time control is around $3.75\times10^{-5}$
after $6.5s$, while states by Nussbaum-gain method is $0.01$. Moreover,
the proposed method has a smaller overshoot and control effort than
Nussbaum-gain method. The reason for this may be that the proposed
method can find a proper control direction quickly by switching logic. 

Next, with the same controller parameters, we consider the control
performance in Cases B-F. Fig. \ref{fig:7} shows the state trajectories
in Cases B and C. We can see that in both cases, the proposed method
has a superior performance with smaller overshoot and faster convergence
rate. Fig. \ref{fig:8} demonstrates the state trajectories in Cases
D and E, it can be seen that control performance of Nussbaum-gain
method deteriorates a lot. For Case F, the Nussbaum-gain method has
become highly unstable. All these show the stronger robustness of
the proposed method. 
\end{example}

%

\begin{example}
	To further verify the validity of the proposed method, an experiment is conducted. The experimental platform is shown in Fig. \ref{fig:ex12}.  It involves the following three main parts: (1) A DC brush motor with an encoder; (2)  A STM32F407 control board. The control board samples the actual velocity every millisecond, which  aims to realize the velocity closed-loop control. (3) A H bridge drive circuit board. We assume that we do not know the control direction. The controller parameters are set as: $K_1=10,K_2=12,\theta_{1}(0)=0.5$
	and for $\sigma_{1}(t)\geq1$, we have $\theta_{1}(\sigma_{1})=(-1)^{\sigma_{1}(t)}(\iota_{11}+\iota_{12}\sigma_{1})$,
	where $\iota_{11}=\iota_{12}=1$. 
	The experimental results are shown in Fig. \ref{fig:ex3}-\ref{fig:ex4}. Fig. \ref{fig:ex3} shows the tracking performance. Fig. \ref{fig:ex4} shows the variations of $V,\eta$ and control input. We can see that the system output can track the reference signal accurately. The control parameters can be adaptively adjusted to identify the system control direction. All these show the validity of the proposed method.
\end{example}



\section{Conclusions}
In this paper, two kinds of logic-based switching adaptive controllers are proposed.
The finite time stability can be guaranteed for the nonlinear
systems suffering from multiple unknown control directions and unstructured
uncertainties.  Future work will be focused on extending the presented
design approach to general hybrid systems. 

\appendices{}

\section{Proof of Lemma \ref{lem:C1}}
\begin{IEEEproof}
	This is a variation of Comparison Principle \cite{key-3}. Subtracting (\ref{eq:C1}) from (\ref{eq:C2}), we have: 
	\begin{equation}
	\dot{e}(t)\leq-a[y^{\gamma}(t)-x^{\gamma}(t)]\label{eq:30-1}
	\end{equation}
	where $e(t)\triangleq y(t)-x(t).$ Then, we only need to prove $e(t)\leq0$
	for $\forall t\in[t_{0},t_{1})$. In fact, if this is not true, there
	must exist a time instant $t^{*}\in[t_{0},+\infty)$ such that $e(t^{*})>0$.
	Since $e(t_{0})=y(t_{0})-y(t_{0})-\varepsilon=-\varepsilon\leq0$,
	we can conclude that there exists a time interval $[\tau_{1},\tau_{2}]\subseteq[t_{0},t^{*}]$
	such that $e(\tau_{1})=0$, $e(t)>0$ for $\forall t\in(\tau_{1},\tau_{2}]$.
	By Mean Value Theorem, there exists a time instant $\tau_{3}\in[\tau_{1},\tau_{2}]$
	such that $e(\tau_{3})>0$ and $\dot{e}(\tau_{3})>0$. However, this
	contradicts (\ref{eq:30-1}). It completes the proof. 
\end{IEEEproof}

\section{Proof of Claim 1a}

The proof is divided into two parts. For the first part, we will show
Claim 1a holds with $m=0$. The second part will show Claim 1a is
true for any finite integer $m$. 

\textbf{Part I }

We will prove the following claim.\textbf{\smallskip{}
}

\noindent \textbf{Claim 1a$\mathbf{'}$.} \label{prop:1a-1}\textit{Suppose
	$|s_{i}(t_{s}^{0})|<\hat{\chi}_{i}(t_{s}^{0})$ and $\eta_{i}(t_{s}^{0})>V_{i}(\overline{x}_{i}(t_{s}^{0}),t_{s}^{0})$
	for $\forall i=1,2,...,n$, then we have}

\textit{1) The considered closed loop nonlinear system admits continuous
	solution $\overline{x}_{n}(t)$ on $[t_{s}^{0},t_{s}^{1})$; }

\textit{2) There exists a positive constant $\delta_{i}^{0}$ such
	that $|s_{i}(t)|\leq\hat{\chi}_{i}(t_{s}^{0})-\delta_{i}^{0}$ for
	$\forall i=1,2,...,n$ on $[t_{s}^{0},t_{s}^{1})$; }

\textit{3) All the signals in the system are bounded on $[t_{s}^{0},t_{s}^{1})$.}
\begin{IEEEproof}
	The proof is divided into following phases.
	
	1) Prove that the closed loop system admits continuous solution $\overline{x}_{n}(t)$ on
	$[t_{s}^{0},t^{*})$ with $|s_{i}(t)|<\hat{\chi}_{i}(t_{s}^{0})(i=1,2,...,n)$
	and $t_{s}^{0}<t^{*}\leq t_{s}^{1}$.
	
	According to the Initialization in Algorithm 1, we know $|s_{i}(t_{s}^{0})|<\hat{\chi}_{i}(t_{s}^{0})$
	and $\eta_{i}(t_{s}^{0})>V_{i}(\overline{x}_{i}(t_{s}^{0}),t_{s}^{0})$.
	Meanwhile, nonlinear functions $h_{i}(\overline{x}_{i}),f_{i}(\overline{x}_{i})$
	in (\ref{eq:1-1}) are continuously differentiable and adaptive parameters
	$\hat{\chi}_{i},\hat{\Theta}_{i}$ are constants on $[t_{s}^{0},t_{s}^{1})$.
	Then, according to \cite{key-20a} and \cite{key-s}, we know the above statement is true.
	
	2) Prove $t^{*}$ can be extended to $t_{s}^{1}$. That is the closed
	loop system admits continuous solution $\overline{x}_{n}(t)$ on $[t_{s}^{0},t_{s}^{1})$
	with $|s_{i}(t)|<\hat{\chi}_{i}(t_{s}^{0})(i=1,2,...,n)$. 
	
	This is proved by contradiction. If this is not true, then $t^{*}<t_{s}^{1}$
	and there exists an integer $i'\in\{1,2,...,n\}$ such that as $t\rightarrow t^{*}$,
	$|s_{i'}(t)|\rightarrow\hat{\chi}_{i'}(t_{s}^{0})$. Meanwhile, $|s_{i}(t)|<\hat{\chi}_{i}(t_{s}^{0})$
	for $\forall i=1,2,...,n$ on $[t_{s}^{0},t^{*})$. 
	
	Note that from (\ref{eq:33}), we know $\dot{\overline{\eta}}\leq-a'\overline{\eta}^{\frac{1+\alpha}{2}}\leq0$
	where we have used the fact that $\eta_{i}\geq V_{i}\geq0(i=1,...,n)$
	on \textit{$[t_{s}^{0},t^{*})$.} This means that all the Lyapunov
	functions $V_{i}(i=1,...,n)$ are bounded. Next, analysis will be taken on each
	step in the controller design to seek a contradiction.
	
	\textit{Step} \textit{1}. Since $V_{1}$ is bounded, from (\ref{eq:b1})
	we conclude that there exists a positive constant $\delta_{1}^{0}$
	such that $|s_{1}|\leq\hat{\chi}_{1}(t_{s}^{0})-\delta_{1}^{0}$.
	Using (\ref{eq:n2}) and (\ref{eq:n3}), we know $x_{2}^{*},x_{2}$
	are both bounded.
	
	\textit{Step i}(\textbf{$2\leq i\leq n$}). Due to $x_{i}^{*},V_{i}$
	are bounded, from Proposition \ref{prop:9-1-1} we know there exists
	a positive constant $\delta_{i}^{0}$ such that $|s_{i}|\leq\hat{\chi}_{i}(t_{s}^{0})-\delta_{i}^{0}$.
	Using (\ref{eq:n2})-(\ref{eq:n4}), we conclude that $x_{i+1}^{*},x_{i+1}$
	are both bounded with $x_{n+1}\triangleq0$.
	
	Therefore, we have $|s_{i}|\leq\hat{\chi}_{i}(t_{s}^{0})-\delta_{i}^{0}$
	for $i=1,2,...,n$. This contradicts the fact that there exists an
	integer $i'\in\{1,2,...,n\}$ such that when $t\rightarrow t^{*}$,
	$|s_{i'}(t)|\rightarrow\hat{\chi}_{i'}(t_{s}^{0})$. Then, we can
	conclude that the system has continuous solution on $[t_{s}^{0},t_{s}^{1})$
	with $|s_{i}(t)|<\hat{\chi}_{i}(t_{s}^{0})$. 
	
	3) Prove Claim 1a$'$ is true.
	
	This can be proved by following the line of the above procedures and
	using the fact that $|s_{i}(t)|<\hat{\chi}_{i}(t_{s}^{0})(i=1,2,...,n)$
	on $[t_{s}^{0},t_{s}^{1})$. 
\end{IEEEproof}
\textbf{Part II }

We will prove Claim 1a holds for any finite integer $m$.

According to Claim 1a$'$, we know the system admits continuous solution
on $[t_{s}^{0},t_{s}^{1})$. Then, by Algorithm 1 and Remark \ref{rem:(Idea-of-Algorithm},
it is not hard to show at switching time $t_{s}^{1}$, we have

1) $s_{i}(t_{s}^{1}),\hat{\chi}_{i}(t_{s}^{1}),\eta_{i}(t_{s}^{1}),V_{i}(\overline{x}_{i}(t_{s}^{1}),t_{s}^{1})$
are all bounded for $\forall i=1,2,...,n$;

2) $|s_{i}(t_{s}^{1})|<|\hat{\chi}_{i}(t_{s}^{1})|$ and $\eta_{i}(t_{s}^{1})>V_{i}(\overline{x}_{i}(t_{s}^{1}),t_{s}^{1})$
for $\forall i=1,2,...,n$. 

Then, by regarding $t_{s}^{1}$ as a new initial time and repeating
the procedures in Claim 1a$'$, we can prove Claim 1a holds with $m=1$.
Similarily, we can prove the result for any finite $m$. The proof
is completed.

\end{document}